\documentclass[a4paper,10pt]{article}%

\usepackage{geometry}
\usepackage{amsmath}
\usepackage{amsfonts}
\usepackage{amssymb}
\usepackage{graphicx}
\usepackage{indentfirst}
\usepackage{bbold}
\usepackage[small,bf,hang]{caption}
\usepackage{slashed}
\usepackage{braket}
\usepackage[pdftitle={Proceed_WWNQFT},colorlinks=true]{hyperref}

\usepackage[T1,T5]{fontenc}

\providecommand{\U}[1]{\protect\rule{.1in}{.1in}}

\renewcommand{\d}{\ensuremath{\mathrm{d}}}
\newcommand{\ii}{\ensuremath{\mathrm{i}}}
\newcommand{\p}{\partial}

\newcommand{\e}{\ensuremath{\mathrm{e}}}

\newcommand{\R}{\ensuremath{\mathrm{R}}}
\newcommand{\YM}{\ensuremath{\mathrm{YM}}}
\newcommand{\FP}{\ensuremath{\mathrm{FP}}}
\newcommand{\gf}{\ensuremath{\mathrm{gf}}}

\newcommand{\MSbar}{\overline{\mbox{MS}}}

\newcommand{\lms}{\Lambda_{\overline{\mbox{\tiny{MS}}}}}

\newcommand{\omu}{\overline{\mu}}

\begin{document}

\date{}
\title{{\bf Confinement interpretation in a Yang-Mills $+$ Higgs theory when considering Gribov's ambiguity}}
\author{
\textbf{I.~F.~Justo }$^{ab}$\thanks{igorfjusto@gmail.com}\,\,,
\textbf{M.~A.~L.~Capri}$^{a}$\thanks{caprimarcio@gmail.com}\,\,,
\textbf{D.~Dudal}$^{bc}$\thanks{david.dudal@kuleuven-kulak.be}\,\,,
\textbf{A.~J.~G\'{o}mez}$^{a}$\thanks{ajgomez@uerj.br}\,\,,
\textbf{M.~S.~Guimaraes}$^{a}$\thanks{msguimaraes@uerj.br}\,\,,
\\
\textbf{S.~P.~Sorella}$^{a}$\thanks{sorella@uerj.br}\,\,, 
\textbf{D.~Vercauteren}$^{d}$\thanks{vercauterendavid@dtu.edu.vn}\, 
\thanks{Work supported by FAPERJ, Funda{\c{c}}{\~{a}}o de Amparo {\`{a}} Pesquisa do Estado do Rio de Janeiro, under the program \textit{Cientista do Nosso Estado}, E-26/101.578/2010.}\,\,
\\[2mm]
{\small \textnormal{$^{a}$  \it Departamento de F\'{\i }sica Te\'{o}rica, Instituto de F\'{\i }sica, UERJ - Universidade do Estado do Rio de Janeiro,}}
\\ 
{\small \textnormal{\phantom{$^{a}$} \it Rua S\~{a}o Francisco Xavier 524, 20550-013 Maracan\~{a}, Rio de Janeiro, Brasil}}
\\
{\small \textnormal{$^{b}$ \it Ghent University, Department of Physics and Astronomy, Krijgslaan 281-S9, 9000 Gent, Belgium}}
\\
{\small  \textnormal{$^{c}$ \it KU Leuven Campus Kortrijk - KULAK, Department of Physics, Etienne Sabbelaan 53, 8500 Kortrijk, Belgium}}
\\
{\small \textnormal{$^{d}$ \it Duy T\^an University, Institute of Research and Development, P809, K7/25 Quang Trung, {\fontencoding{T5}\selectfont H\h ai Ch\^au, \DJ\`a N\~\abreve ng}, Vietnam}}
\normalsize
}

\maketitle

\begin{abstract}
This work presents concisely the results obtained from the analysis of the two-point function of the gauge field in the $SU(2)$ and $SU(2)\times U(1)$ gauge theories, in the Landau gauge, coupled to a scalar Higgs field in the fundamental or adjoint representation. Non-perturbative effects are considered by taking into account the Gribov ambiguity. In general, in both Yang-Mills models the gluon propagator has non-trivial contributions of physical and non-physical modes, which clearly depends on the group representation of the Higgs field. These results were presented during the $4^{th}$ Winter Workshop on Non-perturbative Quantum Field Theory, which took place in Sophia-Antipolis - France.
\end{abstract}

%%%%%%%%%%%%%%%%%%%%%%%%%%%%%%%%%%%%%%%%%%%%%%%%%
\section{Introduction}
The understanding of non-perturbative aspects of non-Abelian gauge theories is one of the main challenging problems in quantum field theories. As an example, we may quote the transition between the confined and the Higgs regime in a Yang-Mills theory coupled to a scalar Higgs fields. See refs.\cite{Polyakov:1976fu,Cornwall:1998pt,Baulieu:2001vw} for analytical investigations and  \cite{Fradkin:1978dv,Nadkarni:1989na,Hart:1996ac,Caudy:2007sf,Maas:2010nc,Greensite:2011zz} for results obtained through numerical lattice simulations.

%\footnote{By Higgs regime we mean the region in the parameter space where the Higgs mechanism takes place, and where the effects of Gribov's ambiguities are not sensitive. For confining regime we mean the region where non-perturbative effects become relevant leading to a modified gluon propagator, compared with that one in the standard Higgs mechanism.}

Non-perturbative effects can be accounted perturbatively by considering ambiguities in the gauge fixing process, first noted by Gribov in \cite{Gribov:1977wm}. These ambiguities, also referred to as Gribov copies, are unavoidably present in the Landau gauge, since the (Hermitian) Faddeev-Popov operator admits the existence of zero modes. A brief introduction to the Gribov issue is presented in section \ref{introductiontogribov}, based on the works \cite{Gribov:1977wm,Sobreiro:2005ec,Vandersickel:2012tz}. With that being said, the analysis of Yang-Mills theories coupled to a scalar Higgs field, in its fundamental and adjoint representation, while taking into account non-perturbative effects arising from the Gribov issue, becomes of great aid in the comprehension of the confined/Higgs transition effect. More specifically, the $SU(2)$ and $SU(2) \times U(1)$ Yang-Mills models are analysed in this work, in the $3$-- and $4$--dimensional cases. Some general features and definitions concerning the Yang-Mills $+$ Higgs field are worked out in section \ref{The Yang-Mills $+$ Higgs field theory}.

Part of the dynamics of the system can be read off from the structure of the gluon propagator, which is a function of the parameter space of each model and reflects the existence of Gribov's issue. The Higgs and confinement regimes are detected as follows: when the gauge field displays a Yukawa type propagator the system is said to be in the {\it Higgs regime}; for a propagator of the Gribov type the system is in the {\it confined regime}. The confinement interpretation follows from the Gribov propagator that lacks the usual physical particle interpretation, since it exhibits complex conjugate poles. As such, gluons do not belong to the physical spectrum of the theory, which is compatible with the effect of confinement \cite{Baulieu:2009ha,Sorella:2010it}. Note that the ``phase structure'' in each theory turns out to be dependent on the group representation of the Higgs field. The analysis of each model, $SU(2)$ and $SU(2)\times U(1)$, are carried out in sections \ref{gfwithbroksym} and \ref{The Electroweak theory}, respectively. In the last section \ref{conclusion} our conclusions are presented.

The content of the present work is a summary of the results found by the authors in the works \cite{Capri:2013oja,Capri:2013gha,Capri:2012ah,Capri:2012cr}.

\section{A brief introduction to Gribov's issue}
\label{introductiontogribov}
%%%%%%%%%%%%%%%%%%%%%%%%%%%%%%%%%%%%%%%%%%%%%%%%%%%%%%%%%%%%%%%%%%%%
%{\bf Rewrite!!}\\
%The procedure of gauge fixing and its requirement for a proper quantization of a gauge field are well known issues in quantum field theory. This subject is treated in different ways by many text books  (see \cite{Weinberg:1996kr,Peskin:1995ev,Srednicki:2007qs} for a few examples) and this work does not intend to be one more different approach. On the contrary, we aim to provide a careful analysis on the Faddeev-Popov's procedure on gauge fixing, in the path integral formalism, pointing out a few crucial steps, which are not always made clear in standard text books. Once this process is done Gribov's problem arises as a necessity to properly manage those important points, in order to rigorously ensure the gauge fixing condition. Gribov was the first one to point out and to propose a first consistent approach to fix some inconsistencies that arise when the Lorentz gauge condition --- specifically, the Landau one --- is chosen \cite{Gribov:1977wm}. Since then, the inconsistencies, known as Girbov's ambiguities, became subject of intense analytic and numerical studies as long as it provides a portal to nonperturbative physics and possibly sign of confinement.

This section is organized in a way to provide a brief introduction to Gribov's ambiguities by following his seminal work \cite{Gribov:1977wm}. We do not intend to provide the final word on this matter and as such we would refer to \cite{Vandersickel:2012tz} for a more complete and pedagogical reference. The Faddeev-Popov quantization procedure is reproduced (not in full detail) in the subsection \ref{introductiontogribov1}, while Gribov's problem will be introduced and implemented on the path integral in the two subsequent subsections, \ref{Gribov's issue} and \ref{Gribovimplementation}. In the last subsection \ref{A sign of confinement from gluon propagator}, we should present one of the most important outcomes of the Gribov's issue: a possible interpretation of gluon confinement, which is encoded in the poles of the gluon propagators. We should say, to clarify matters, that our present work concerns computations up to one-loop in perturbation theory.

\subsection{Quantization of non-Abelian gauge field}
\label{introductiontogribov1}

\begin{figure}[h]
\begin{center}
\includegraphics[width=8cm]{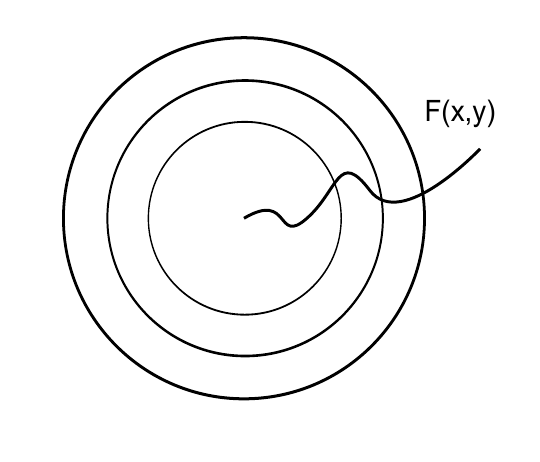}
\caption{Gauge orbits of a system with rotational symmetry in a plane and a function $\mathcal F$ which picks one representative from each gauge orbit.}
\label{fig1}
\end{center}
\end{figure}

The gauge invariant action of a non-Abelian gauge field, or the Yang-Mills ($\YM$) action, is given by
\begin{eqnarray}
S_{\text{YM}} &=& \int \mathrm{d}^d x \frac{1}{4}  F_{\mu\nu}^a  F_{\mu\nu}^a \;,
\label{YMact}
\end{eqnarray}
with $
F^{a}_{\mu\nu} ~=~ \partial_{\mu}A^{a}_{\nu} - \partial_{\nu}A^{a}_{\mu} + gf^{abc}A^{b}_{\mu}A^{c}_{\nu}
$ 
being the field strength tensor. The action \eqref{YMact} enjoys the feature of being symmetric under gauge transformation, which is defined for the gauge field as
\begin{eqnarray}
A_{\mu}' ~=~ U^{\dagger}A_{\mu}U - \frac{\ii}{g} U^{\dagger}(\p_\mu U)\;,
\label{gaugetransf}
\end{eqnarray}
with $U(x) \in SU(N)$ and $N$ being the number of colours. A geometrical representation of this symmetry can be seen in Figure \ref{fig1}, where each orbit, representing equivalent gauge fields, is crossed by a gauge curve ${\cal F}$. It means that, the YM action is invariant under transformations that keep the transformed gauge field on the same gauge orbit.

As is well known, that gauge invariance of the action induces inconsistencies in the quantization of the gauge field reflecting the existence of infinite physically equivalent configurations. To get rid of those spurious configurations from the system one has to fix the gauge, which is, in the geometrical view, to choose a convenient curve ${\cal F}$ that crosses only once each gauge orbit. In the path integral formalism, the gauge fixing procedure is carried out by the Faddeev-Popov procedure.

The (Euclidean) gauge fixed partition function reads
\begin{eqnarray}
\label{genym}
Z_{\text{YM}}(J) ~=~ \int_{\cal F} [\d A]\;\; \e^{-S_{\text{YM}} + \int {\rm d} x J_\mu^a A_\mu^a}\;,
\label{YMaction0}
\end{eqnarray}
where $\int_{\cal F}$ denotes the path integral restricted to the curve $\cal F$, which can be recast in the form

\begin{eqnarray}
Z_{\YM}(J) ~=~ V\int [\d A] \;\; \Delta_{\cal F}(A) \delta({\cal F}(A)) \mathrm{e}^{-S_{\text{YM}} + \int {\rm d} x J_\mu^a A_\mu^a}\;.
\label{genfp}
\end{eqnarray}
The $V$ factor accounts for the (infinite) orbit's volume, while $\Delta_{\cal F}(A)$ stands for the Jacobian of infinitesimal gauge transformations 
\begin{eqnarray}
A'_{\mu} ~=~ A_{\mu} - D_{\mu}\theta(x) \;,
\label{inftgaugetransf}
\end{eqnarray}
which is needed since we are working with the gauge transformed integration measure. The $\theta(x) ~=~ \theta^{a}(x)\tau^{a}$ stands for the infinitesimal gauge transformation parameter, while $\tau^{a}$ are the $SU(N)$ generators.  $\delta({\cal F}(A))$ is the delta function ensuring the gauge condition $\cal F(A)=0$. We should emphasize that the Jacobian of a given transformation (as the infinitesimal gauge transformation) is defined as the absolute value of the determinant of the derivative --- with respect to the transformation parameter ($\theta$) ---  of the transformed field. In our gauge fixing case we have
\begin{equation}
\Delta_{\mathcal F} (A) ~=~  |\det \mathcal  M_{ab} (x,y) | \qquad \text{with} \qquad  \mathcal M_{ab} (x,y) ~=~ \left. \frac{\delta \mathcal F^a (A'_\mu (x))  }{\delta \theta^b(y)} \right|_{\mathcal F(A') =0} \;,
\label{absvalue}
\end{equation}
with 
%At this point we should chose the gauge fixing condition ${\cal F}(A)$. The chosen one will be the family of gauges defined by the Lorentz condition, namely
\begin{eqnarray}
\mathcal{F}^a (A'_\mu (x)) ~=~ \p_{\mu} A_{\mu }^{\prime a} (x) - \xi b^a (x) \;,
\label{notideal}
\end{eqnarray}
being the gauge condition and $A'_{\mu}$ given by \eqref{inftgaugetransf}. $b^{a}$ stands for the Lagrange multiplier: an arbitrary scalar field (the Nakanishi-Lautrup field).  With this we have the Faddeev-Popov operator $ \mathcal M_{ab} (x,y) $, which reads
\begin{eqnarray}\label{mab}
 \mathcal M_{ab} (x,y) ~=~  \left.    - \p_\mu  D_\mu^{ab} \delta (y-x)    \right|_{ \mathcal F(A) = 0 } \;.
\label{fpoperator}
\end{eqnarray}
Following the standard procedure by Faddeev-Popov to implement the gauge fixing, one ends up with
\begin{eqnarray}
Z(J) &=&   \int [\d A][\d c] [\d \overline c]         \exp \left[- S_\YM + \underbrace{\int \d x \left(   \overline c^a   \p_\mu  D_\mu^{ab}  c^b -  \frac{1}{2 \xi} (\p_\mu A_\mu^a )^2   \right)}_{S_\gf} + \int \d x J_\mu^a A_\mu^a  \right]\;,
\label{genfuncfp}
\end{eqnarray}
where $S_{\gf}$ stands for the gauge fixing term. The ghost fields $(c,\bar{c})$ are anti-commuting fields with ghost number $+1$ and $-1$, respectively. At the end we have the Faddeev-Popov action given by $S_{\FP} = S_{\YM} + S_{\gf}$.

%The implementation of the functional determinant in the action should be done through its Gaussian functional integration representation,
%\begin{eqnarray}
%\det \mathcal M_{ab} (x,y) &=& \int [\d c] [\d \overline c] \exp \int \d x \d y \overline c^a (x) \mathcal M_{ab} (x,y) c^b (y)\;,
%\label{absvalue2}
%\end{eqnarray}
%by the insertion of a couple of anti-commuting fields $(c^{a},\,\overline{c}^{a})$, known as the Faddeev-Popov ghosts, carrying anti-commuting charge number (ghost number) $+1$ and $-1$, respectively. Now, there remains the delta function to be implemented in the action. Since the partition function \eqref{genfp} does not depend on the auxiliary field, one could add a unitary and normalized weight to the delta function, namely
%\begin{eqnarray}
%  \int [\d b] \delta (\p A - \xi b ) \exp \left( \frac{\xi}{2} \int \d x \;\; b^{a}b^{a} \right)  &=& \exp \left( \frac{1}{2 \xi} \int \d x (\p_\mu A_\mu^a )^2 \right)  \;.
%\end{eqnarray}
%Note that when $\xi \to 0$ the Gaussian oscillates very rapidly acting like a delta function, imposing the Landau gauge $\p A = 0$. Besides that, for any non-zero value of $\xi$ the Gaussian integral does not represent a delta function and, thus, no gauge condition is imposed over the gauge field anymore. Finally, one could write down the whole expression of the gauge fixed action as

%with the following generating functional,

Let us emphasize two important assumptions made in the process to obtain \eqref{genfuncfp}:
\begin{itemize}
\item The gauge condition ${\cal F}^{a}$ is said to pick up only one field configuration from each gauge orbit, representing the physical equivalent configurations related by gauge transformations;

\item The determinant of $\mathcal M_{ab} (x,y)$, present in the Jacobian definition \eqref{absvalue}, is supposed to be always positive.
\end{itemize}
Gribov showed in \cite{Gribov:1977wm} that those assumptions are not always true. Furthermore, he showed that the failure of the first statement is closely related to the failure of the second one and, besides that, he pointed out the existence of zero mode configurations related to the $\FP$ operator \eqref{fpoperator}. Those subtleties give rise to the Gribov issue, summarized in the next subsection.

\subsection{Gribov's issue}
\label{Gribov's issue}

\begin{figure}[h]
\begin{center}
\includegraphics[width=8cm]{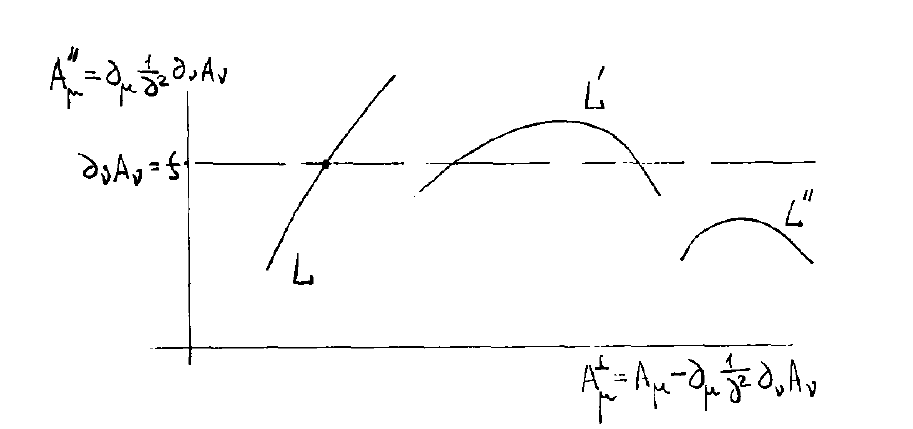}
\caption{The gauge condition curve can cross each orbit of equivalence once, more then once and at no point at all. The horizontal axes denotes the transversal gluon propagator, while the vertical axis represents the longitudinal one \cite{Gribov:1977wm}.}
\label{2fig1}
\end{center}
\end{figure}

As stated before, one of the most important hypotheses is that once the gauge fixing condition is chosen, ${\cal F}=0$, for a given gauge field configuration $A'_{\mu}$, one should be able to find out only one gauge field configuration $A_{\mu}$ related to $A'_{\mu}$ through a gauge transformation that fulfils the gauge condition. This situation is represented in Figure \ref{2fig1} by the curve $\bf L$. It means that we cannot find gauge-equivalent configurations, related by gauge transformation, satisfying the gauge condition \eqref{notideal}, which is represented by the curve $\bf L'$. Besides that, we cannot find a situation where no one field configuration on a gauge orbit satisfy the gauge condition \eqref{notideal}, which is represented by the curve $\bf L''$.

While there is no known example where the curve $\bf L''$ does exist, the situation described by curve the $\bf L'$ is quite typical in non-Abelian gauge theories and, therefore, it is worth analysing \cite{Gribov:1977wm,Sobreiro:2005ec,Vandersickel:2012tz}. To this end, let us consider two close gauge-equivalent configurations: $A'_{\mu}$ and $A_{\mu}$ related by an infinitesimal gauge transformation \eqref{inftgaugetransf} and satisfying the Landau gauge condition (covered for $\xi \to 0$). That is,
\begin{eqnarray}
&A_\mu' ~=~ U A_\mu U^{\dagger} - \frac{\ii}{g} (\p_\mu U) U^{\dagger}\;,  \qquad  \p_\mu A_\mu ~=~ 0 \quad \& \quad  \p_\mu A_\mu' ~=~ 0 \;.
\end{eqnarray}
Then, we are led to
\begin{eqnarray}
- \p_\mu D_\mu \alpha &=& 0\;,
\label{FPeigenvlueqtion}
\end{eqnarray}
with the following covariant derivative $D_{\mu}^{ab} ~=~ \p_{\mu}\delta^{ab} - gf^{abc}A^{c}_{\mu}$. Equation \eqref{FPeigenvlueqtion} does point to the existence of zero eigenvalues of the $\FP$ operator. The conclusion is that if there are (at least) two close equivalent fields satisfying the Landau gauge, defining gauge copies, then the Faddeev-Popov operator has zero modes (eigenstates associated to zero eigenvalues). Note that for the Abelian case the equation reduces to the Laplace equation,
\begin{eqnarray}
\p^{2} \alpha &=& 0\;.
\label{abelianeingvalue}
\end{eqnarray}
Since it defines plane waves, thus not normalisable, we are not considering this case, restricting ourselves to the analysis of fields that vanish at infinity. It is clear one needs to study the $\FP$ operator's eigenvalue. To that end, suppose the follow eigenvalue equation
\begin{eqnarray}
- \p_\mu D_\mu \psi ~=~ \epsilon \psi\;.
\end{eqnarray}
Clearly, if $A_{\mu}$ is small enough then we have 
\begin{equation}
-\p_\mu^2 \psi = \epsilon \psi \;,
\end{equation}
which is solvable only for positive $\epsilon$, since in the momentum space we have $p^{2} = \epsilon >0$. Then, for small enough field configurations we do not have zero mode issues anymore. Otherwise, if $A_{\mu}$ turns out to be large, but not too large, we reach the solution $\epsilon = 0$, showing the existence of zero modes. For even large amplitudes of $A_{\mu}$ the eigenvalue turns to be negative; if it keeps growing then $\epsilon$ reaches zero again, and so on.

In his work \cite{Gribov:1977wm} Gribov showed that the domain of functional integration should be restricted to the first region, named ``first Gribov region'' $\Omega$, where $\epsilon >0$, in order to avoid gauge copies. It is known that this region is actually not free of copies. It has being analytically motivated that the existence of those copies inside $\Omega$ do not influence physical results (see \cite{Dudal:2014rxa} and references therein). Furthermore, up to now there is no way to implement analytically the restriction of the partition function to the region free of copies (known as Fundamental Modular Region, $\Lambda$).

%(see Figure \ref{2fighorizon})

%\begin{figure}[h]
%\begin{center}
%\includegraphics[width=8cm]{gribovhorizon.pdf}
%\caption{The different regions in the hyperspace $\p A = 0$.}
%\label{2fighorizon}
%\end{center}
%\end{figure}

%(see Figure \ref{2fighorizon})

The first Gribov region $\Omega$ enjoys some useful properties that have mathematical proof only in the Landau gauge, which justifies our interest in this gauge (see \cite{Vandersickel:2012tz} and references therein). Namely,
\begin{itemize}
\item 
For every field configuration infinitesimally close to the border $\delta\Omega$  and belonging to the region immediately out side $\Omega$ (called $\Omega_{2}$), there exist a gauge-equivalent configuration belonging to $\Omega$ and infinitesimally close to the border $\delta\Omega$ as well \cite{Gribov:1977wm}. It was also proven that every gauge orbit intersects the first Gribov region $\Omega$ \cite{Capri:2005dy,Zwanziger:1982na}.

\item
The Gribov region is convex \cite{Zwanziger:2003cf}. This means that for two gluon fields $A_\mu^1$ and $A_\mu^2$ belonging to the Gribov region, also the gluon field $A_\mu = \alpha A_\mu^1 + \beta A_\mu^2$ with $\alpha, \beta \geq 0$ and $\alpha + \beta  =1$, is inside the Gribov region.

\item
One may also show that the Gribov region is bounded in every direction \cite{Zwanziger:2003cf}.
\end{itemize}

For more details concerning (proofs of) properties of Gribov regions see \cite{Vandersickel:2012tz} and references therein.

\subsection{Implementing the restriction to the first Gribov region $\Omega$}
\label{Gribovimplementation}
As was discussed before, our aim is to restrict the domain of the functional integration to the region where the $\FP$ operator has positive eigenvalues. Therefore, let us define the first Gribov region as the region where the $\FP$ operator is positive definite. Namely,
\begin{eqnarray}
\Omega ~\equiv ~ \{ A^a_{\mu}, \, \p_{\mu} A^a_{\mu} ~=~  0, \, \mathcal{M}^{ab}  >0  \} \;,
\label{defgribovregion}
\end{eqnarray}
where $\mathcal{M}^{ab}$ is the FP operator,
\begin{eqnarray}\label{mab2}
\mathcal M^{ab}(x,y) ~=~ - \p_\mu  D_\mu^{ab} \delta(x-y)  ~=~ \left( - \p_\mu^2 \delta^{ab} +  \p_\mu f_{abc} A_\mu^c \right) \delta(x-y)\;,
\end{eqnarray}
and the condition on this operator defines a positive definite operator, given by
\begin{eqnarray}\label{positivedef}
\int \d x \d y \omega^a(x)  \mathcal M^{ab} (x,y) \omega^b(y) > 0 \;.
\end{eqnarray}
Since the $\FP$ operator is the inverse of the ghost propagator, the ghost propagator plays a central role. With this being said, we are going to compute the ghost propagator till the first loop order, by following \cite{Gribov:1977wm}. The effective implementation of the restriction of the functional integration will be done by imposing the no-pole condition, which will become clear in what follows. The computation of the ghost propagator can be done by making use of the partition function \eqref{genfuncfp}. Let us denote the ghost propagator as
\begin{eqnarray}
\left\langle  c^{a}(p) \overline{c}^{b}(-p)  \right\rangle ~=~ \mathcal G (k^2, A)_{ab}\;.
\end{eqnarray}
Treating the gauge field as a classical external field we get
\begin{eqnarray}
\mathcal G (k^2, A) &=&  \frac{1}{N^2 -1}\delta_{ab} \mathcal G (k^2, A)_{ab} ~=~  \frac{1}{k^2} + \frac{1}{V}\frac{1}{k^4} \frac{N g^2}{N^2 - 1} \int\frac{ \d^d q}{(2 \pi)^d} A_\mu^\ell(-q) A_\nu^{\ell} (q)  \frac{(k-q)_\mu  q_\nu}{(k-q)^2} \nonumber\\
&=& \frac{1}{k^2}\left( 1 + \sigma(k, A) \right)\;,
\label{coloraway}
\end{eqnarray}
whereby
\begin{eqnarray}
\sigma(k,A) ~=~ \frac{1}{V}\frac{1}{k^2} \frac{Ng^2}{N^2 - 1} \int\frac{ \d^d q}{(2 \pi)^d} A_\mu^\ell(-q) A_\nu^{\ell} (q) \frac{ (k-q)_\mu  k_\nu }{(k-q)^2}\;.
\label{1ghostformfacto}
\end{eqnarray}
The ghost propagator \eqref{coloraway} can be perturbatively approximated by

%Note that the first correction in Feynman diagrams (see Figure \ref{ghost}) is not actually the first non-null contribution. By momentum conservation, it is not difficult to see that this first contribution should be null. Therefore, one has to compute the zero order diagram and the second order diagram as well.

%\begin{figure}[h]
%\begin{center}
%\includegraphics[width=16cm]{ghost.pdf}
%\caption{The ghost propagator with external field to second order.}
%\label{ghost}
%\end{center}
%\end{figure}

\begin{eqnarray}
\mathcal G (k^2, A) ~\approx~  \frac{1}{k^2}\frac{1}{( 1 - \sigma(k, A) )}
\label{ghstprop00}
\end{eqnarray}
whereby we can see that for $ \sigma(k, A) < 1$ the domain of integration is safely restricted to $\Omega$, characterising the no-pole condition. It can be proved that $\sigma(k,A)$ is a decreasing function of $k$, which means that the largest value of $\sigma$ is obtained at $k=0$. Thus, if the condition $ \sigma(0, A) < 1$ is ensured then it is completely safe for any non zero value of $k$. Note that the only allowed pole is at $k^2=0$, which has the meaning of approaching the boundary of the region $\Omega$. The partition function restricted to $\Omega$ then becomes,
\begin{eqnarray}\label{ZJ}
Z(J) &=& \int_\Omega [\d A]   \exp \left[- S_\FP   \right] 
\nonumber\\
&=& \int [\d A][\d c] [\d \overline c]    V(\Omega)     \exp \left[- S_\YM  -   \int \d x \left(  \overline c^a   \p_\mu  D_\mu^{ab}  c^b -  \frac{1}{2 \xi} (\p_\mu A_\mu^a )^2  \right)  \right] \;,
\end{eqnarray}
with
\begin{eqnarray}
V(\Omega) &=& \theta (1 - \sigma(0,A))\;,
\label{nopolestepfunct}
\end{eqnarray}
where $\theta (1 - \sigma(0,A))$ is the Heaviside step function ensuring the no-pole  condition. Considering the transversality of the gauge field in the Landau gauge and by making use of the integral representation of the Heaviside step function, one gets the following expression of the partition function restricted to the first Gribov region $\Omega$ \footnote{For more details concerning the derivation of the partition  function restricted to the first Gribov region take a look at \cite{Gribov:1977wm,Sobreiro:2005ec,Vandersickel:2012tz}.},
\begin{equation}
Z(J) ~=~  \mathcal N \int \frac{\d \beta}{2\pi \ii \beta} \int {\cal D} A    \e^{\beta (1 - \sigma(0,A))}    \e^{ - S_\FP }\;,
\label{fullgenertfunct}
\end{equation}
with the ghost form factor given by
\begin{eqnarray}
\sigma(0,A) ~=~
%&=& \frac{1}{V}\frac{1}{d-1} \frac{Ng^2}{N^2 - 1} \lim_{k^2 \to 0} \frac{k_\mu  k_\nu}{k^2} \frac{d-1}{d} \delta_{\mu\nu}  \int\frac{ \d^d q}{(2 \pi)^d} A_\alpha^\ell(-q) A_\alpha^{\ell} (q) \frac{ 1 }{q^2} 
%\nonumber\\
%&=& 
\frac{1}{V} \frac{1}{d} \frac{Ng^2}{N^2 - 1} \int\frac{ \d^d q}{(2 \pi)^4} A_\alpha^\ell(-q) A_\alpha^{\ell} (q) \frac{ 1 }{q^2}\;.
\label{nopole}
\end{eqnarray}
At the end, the final expression for the gauge fixed Yang-Mills action accounting for the Gribov copies reads
\begin{eqnarray}
S\,' &=& S_{\YM} + S_{\gf} +  S_{\beta}\;,
\label{fullgribovact}
\end{eqnarray}
with $S_{\beta} ~=~ \beta\left( \sigma(0,A) - 1\right)$.
As mentioned before, our analytical work shall be restricted to first loop order and therefore we need only to consider quadratic terms of the quantized fields in the action given in equitation \eqref{fullgribovact}. Namely, we have
\begin{equation}
Z_{quad} ~=~ \int \frac{\d\beta e^{\beta }}{2\pi i\beta } [\d A^{a}] \; 
\exp \left\{-\frac{1}{2}  \int \frac{\d^{d}q}{(2\pi )^{d}}  \;  A_{\mu }^{a}(q)\mathcal{P}_{\mu \nu }^{ab }A_{\nu }^{b }(-q)    \right\}  \;,
\label{Zq0}
\end{equation}
with
\begin{eqnarray}
\mathcal{P}_{\mu \nu }^{ab} &=&  \delta^{ab } \left(  q^{2}\delta _{\mu \nu }   +  \left( \frac{1}{\xi } -1 \right) q_{\mu }q_{\nu } +  \frac{2Ng^{2}\beta}{(N^{2}-1)Vd} \frac{\delta _{\mu \nu }}{q^{2}}   \right)   \;.
\label{P0}
\end{eqnarray}
By performing the Gaussian integration of eq.\eqref{Zq0} and making use of the relation
%\begin{equation}
%Z_{quad}  ~=~ \int {\frac{\d \beta}{2\pi i} }
%e^{({\beta} -\ln\beta)} \left(\det\mathcal{P}^{ab}_{\mu\nu}\right)^{-\frac{1}{2}} \;.
%\label{Zq2f00}
%\end{equation}
%That functional determinant may be exponentiated by making use of the relations
\begin{equation}
\left(\det \mathcal{P}_{\mu\nu}^{ab}\right)^{-\frac{1}{2}} ~=~ e^{-\frac{1}{2} \ln \det \mathcal{P}_{\mu\nu}^{ab}} ~=~ e^{-\frac{1}{2}Tr \ln \mathcal{P}_{\mu\nu}^{ab}} \;,
\label{functdeterminant}
\end{equation}
one can finally get
\begin{eqnarray}
Z_{quad} ~=~ \int_{- \ii \infty + \epsilon}^{+ \ii \infty + \epsilon}\frac{\d \beta}{2\pi \ii} \e^{f(\beta)} \;,
\label{ven}
\end{eqnarray}
where
\begin{eqnarray}
f(\beta) ~=~ \beta - \ln \beta  -\frac{d-1}{2} (N^2 - 1) V \int \frac{\d^d q}{(2\pi)^d} \ln \left( q^2 + \frac{\beta N g^2 }{N^2 - 1} \frac{2}{d V}\frac{1}{q^2} \right)\;.
\label{minusfreenergy}
\end{eqnarray}
%which, after taking the trace over all indices of the operator $\mathcal{P}_{\mu\nu}^{ab}$,we have
%\begin{equation}
%\left( \det \mathcal{P}_{\mu \nu }^{ab }\right) ^{-\frac{1}{2}} ~=~ \exp \left[ -\frac{(N^{2}-1)(d-1)V}{2}\int {\frac{\d^{d}q}{(2\pi )^{d}}\ln \left(
%q^{2}   +  \frac{2Ng^2\beta}{(N^{2}-1)dV}   \frac{1}{q^{2}}\right) }\right] \;.
%\label{traceoperator0}
%\end{equation}
The factors $(N^{2}-1)$ and $(d-1)$ in front of the integral came from the trace over the $SU(N)$ gauge group indices and from the trace over the Euclidean space-time indices, respectively\footnote{
A careful computation of the functional trace of $\ln \mathcal{P}_{\mu\nu}^{ab}$ can be found in \cite{Vandersickel:2012tz}}. 
Note that the factor $(d-1)$ is obtained only after the Landau gauge limit is taken.
%\begin{eqnarray}
%Z_{quad} ~=~ \int_{- \ii \infty + \epsilon}^{+ \ii \infty + \epsilon}\frac{\d \beta}{2\pi \ii} \e^{f(\beta)} \;,
%\label{ven}
%\end{eqnarray}
%with
%\begin{eqnarray}
%f(\beta) ~=~ \beta - \ln \beta  -\frac{d-1}{2} (N^2 - 1) V \int \frac{\d^d q}{(2\pi)^d} \ln \left( q^2 + \frac{\beta N g^2 }{N^2 - 1} \frac{2}{d V}\frac{1}{q^2} \right)\;.
%\label{minusfreenergy}
%\end{eqnarray}
In the thermodynamic limit (when $V\to \infty$) the saddle-point approximation becomes exact, so that the integral \eqref{ven} can be easily computed, resulting in
\begin{eqnarray}
\e^{-V{\cal E}_{v}} ~=~  \e^{f(\beta^{\ast})} \;,
\label{relatvaccener}
\end{eqnarray}
with $\beta^*$ solving the equation 
\begin{eqnarray}
\left.
\frac{\partial f}{\partial \beta}\right|_{\beta = \beta^{\ast}}=0\;,
\label{gapeq}
\end{eqnarray}
which lead us to the gap equation
\begin{eqnarray}
\frac{d-1}{d}N g^2    \int \frac{\d^d q}{(2\pi)^d}  \;   \frac{1}{ \left( q^4 + \frac{2\beta^{\ast} N g^2}{(N^2 - 1)dV} \right) }  ~=~ 1   \;.
\label{finalgapeq2}
\end{eqnarray}
%which is called the gap equation. Then, just by deriving equation \eqref{minusfreenergy} we get
%\begin{eqnarray}
%1 -\frac{1}{\beta^{\ast}} - \frac{d-1}{d}N g^2    \int \frac{\d^d q}{(2\pi)^d}  \;   \frac{1}{ \left( q^4 + \frac{2\beta^{\ast} N g^2}{(N^2 - 1)dV} \right) }  ~=~ 0   \;.
%\label{finalgapeq}
%\end{eqnarray}
Note that the Gribov parameter $\beta$, introduced to get rid of gauge ambiguities by restricting the path integral to the first Gribov region $\Omega$, is not a free parameter, it is given dynamically by solving its gap equation \eqref{finalgapeq2}. Besides, it has dimension of $[mass]^{4}$ and is also proportional to the space-time volume $V$. Consequently, in the thermodynamic limit the logarithmic term of eq.\eqref{minusfreenergy} becomes null resulting in eq.\eqref{finalgapeq2}. The function $f(\beta^*)$ can be interpreted, from eq.\eqref{relatvaccener}, as the free-energy of the system. 
%\begin{eqnarray}
%\frac{d-1}{d}N g^2    \int \frac{\d^d q}{(2\pi)^d}  \;   \frac{1}{ \left( q^4 + \frac{2\beta^{\ast} N g^2}{(N^2 - 1)dV} \right) }  ~=~ 1   \;.
%\label{finalgapeq2}
%\end{eqnarray}

%More precisely
%\begin{equation}
%{\cal E}_v = - f(\beta^*)   \;.
% \label{ve}
%\end{equation}
%Interesting enough is that the ghost form factor \eqref{nopole} could be seen as the a perturbative version of the Horizon, computed up to the first loop order in perturbative theory, which is a more sophisticated approach to Gribov's ambiguities \cite{Zwanziger:1989mf}.

\subsection{The gluon propagator}
\label{A sign of confinement from gluon propagator}
In the present subsection we motivate that a possible sign of confinement could be read off from the poles of the gluon propagator, putting this quantity at the centre of any further discussion in the present work. At one-loop order only quadratic terms of the action \eqref{fullgribovact} really matter, so that one can read off the two point function of the gauge field from the inverse of the operator \eqref{P0}, setting $\xi \to 0$ at the very end of the computation. Namely,
\begin{equation}
\left\langle A_{\mu }^{a}(q)A_{\nu}^{b}(-q)\right\rangle ~=~ \frac{q^{2}}{q^{4}  +  \frac{2N g^{2}\beta^{\ast} }{(N^{2}-1)dV} } \left( \delta _{\mu \nu }  -  \frac{q_{\mu }q_{\nu}}{q^{2}}  \right)\delta^{ab}  \;.
\label{Gribovprop0}
\end{equation}
Things become easier to analyze if a new Gribov's parameter is defined,
\begin{eqnarray}
\gamma^{4} = \frac{2\beta^{\ast} N g^2}{(N^2 - 1)dV}\;,
\end{eqnarray}
so that the gluon propagator can be decomposed,
\begin{eqnarray}
\left\langle A_{\mu }^{a}(q)A_{\nu}^{b}(-q)\right\rangle 
%&=& \frac{q^{2}}{q^{4}  +  \gamma^{4} } \left( \delta _{\mu \nu }  -  \frac{q_{\mu }q_{\nu}}{q^{2}}  \right)\delta^{ab} 
%\nonumber \\
&=& \frac{1}{2} \left( \frac{1}{q^{2}  +  \ii \gamma^{2} } +  \frac{1}{q^{2}  -  \ii \gamma^{2} }   \right)  
\left( \delta _{\mu \nu }  -  \frac{q_{\mu }q_{\nu}}{q^{2}}  \right)\delta^{ab} 
 \;.
\label{Gribovprop1}
\end{eqnarray}
Observe from \eqref{Gribovprop1} that the gluon propagator is suppressed in the infrared (IR) regime, while displaying two complex conjugate poles, $m^{2}_{\pm} = \pm \ii\gamma^{2}$. That feature does not allow us to attach the usual physical particle interpretation to the gluon propagator. From the analytic point of view the gluon propagator \eqref{Gribovprop1} has not a(n) (always) positive K\"{a}ll\'en-Lehmann spectral representation, which is necessary to attach a probabilistic interpretation to the propagator\footnote{ See  \cite{Baulieu:2009ha,Sorella:2010it} and references therein for more details on the confinement interpretation of gluons, $i$-particles and the existence of local composite operators, related to these $i$-particles, displaying positive K\"{a}ll\'en-Lehmann spectral representation. For lattice results pointing to the same confinement interpretation see \cite{Cucchieri:2011ig,Cucchieri:2004mf,Cucchieri:2014via}.}. These features lead us to interpret the gluon as being confined.

%%%%%%%%%%%%%%%%%%%%%%%%%%%%%%%%%%%%%%%%%%%%%%%%%%%%%%%%%%%%%%%%%%%%
\section{The Yang-Mills $+$ Higgs field theory}
\label{The Yang-Mills $+$ Higgs field theory}
%%%%%%%%%%%%%%%%%%%%%%%%%%%%%%%%%%%%%%%%%%%%%%%%%%%%%%%%%%%%%%%%%%%%

Before properly starting to analyze the proposed model, let us state a few words on general features that are useful in this work. As mentioned before, the considered $SU(2)$ and $SU(2) \times U(1)$ Yang-Mills gauge theories are coupled to a scalar Higgs field. The Higgs field is considered in either the fundamental and the adjoint representation: in the $SU(2)$ case both the fundamental and the adjoint representation are analysed; while in the $SU(2)\times U(1)$ case only the the fundamental representation will be considered.

Usually the unitary gauge arises as a good choice when the Higgs mechanism is being treated, since in this gauge physical excitations are evident. However, instead of fixing the unitary gauge, we are going to choose the more general $\R_{\xi}$ gauge, whereby the unitary gauge is a special limit, $\xi \to \infty$, and the Landau gauge can be recovered when $\xi \to 0$. In the case of interest, the Landau gauge is imposed at the end of each computation. 
%The unitary gauge is defined by rotating the Higgs scalar field so that this ``rotated field'' becomes orthogonal to the Goldstone modes. It means,
%\begin{eqnarray}
%\tilde{\Phi}_{n} ~=~ \sum_{m}  U_{nm}(x) \Phi_{m}   \;,
%\label{rotatedHfield}
%\end{eqnarray}
%such that
%\begin{eqnarray}
%0   ~=~  \sum_{m,n}\tilde{\Phi}_{n}(x)(\tau^{\alpha})_{nm}\nu_{m}   \;,
%\label{orthogonHfield}
%\end{eqnarray}
%where $\sum_{m}(\tau^{\alpha})_{nm}\nu_{m} $ defines the Goldstone modes ($m,n ~=~ 1,2,...,N$) and $\tau^{a}$ accounts for the $SU(N)$ generators, although here, specifically, the index ``$a$'' runs over the broken symmetry directions. Evidently, the unitary gauge, defined by \eqref{rotatedHfield} and \eqref{orthogonHfield}, is a gauge fixing procedure established in the scalar field, instead of in the vector gauge field by itself.

%Alternatively, instead of impose conditions \eqref{rotatedHfield} and \eqref{orthogonHfield}, we are going to use the condition
The $\R_{\xi}$ gauge condition reads,
\begin{eqnarray}
f^{a} ~=~ \p_{\mu}A^{a}_{\mu} -i\xi  \sum_{m,n} \varphi_{m}(x)(\tau^{a})_{mn} \nu_{n}   \;.
\label{Rxiguage}
\end{eqnarray}
Note that in eq. \eqref{Rxiguage} $i\varphi_{m}(x)(\tau^{a})_{mn} \nu_{n}$ is a possible form for the $b^{a}$ field, defined in \eqref{notideal}. The $\varphi$ field is defined as a small fluctuation around the vacuum configuration of the Higgs field $\Phi$,
\begin{eqnarray}
\Phi(x) ~=~ \varphi(x) + \nu  \;,
\end{eqnarray}
with vacuum expectation value $\langle \varphi \rangle = 0$. In order to apply the gauge condition \eqref{Rxiguage} we should follow the standard process described in many text books, \cite{Weinberg:1996kr,Peskin:1995ev,Srednicki:2007qs}. Rather than a Dirac delta function, as in eq. \eqref{genfp}, one should put a Gaussian function,
\begin{eqnarray}
\delta(f^{a}(A)) \to \exp \left( -\frac{1}{2\xi}\int \d^{d}x\; f^{a}f^{a} \right)   \;.
\label{higgsgauge}
\end{eqnarray}
In the limiting case of $\xi \to 0$ the Gaussian term \eqref{higgsgauge} oscillates very fast around $f^{a} =0$ so that the Gaussian term \eqref{higgsgauge} behaves like a delta function, ensuring the desired Landau gauge. On the other side, if $\xi \to \infty$ then we have the unitary gauge. Needless to say, the limit $\xi \to 0$, recovering the Landau gauge, should be applied at the very end of each computation\footnote{It is perhaps worthwhile pointing out here that the Landau gauge is also a special case of the 't Hooft $R_\xi$ gauges, which have proven their usefulness as being renormalizable and offering a way to get rid of the unwanted propagator mixing between (massive) gauge bosons and associated Goldstone modes, $\sim A_\mu \p_\mu \phi$. The latter terms indeed vanish upon using the gauge field transversality. The upshot of specifically using the Landau gauge is that it allows to take into account potential nonperturbative effects related to the gauge copy ambiguity.}.

In the present work we should deal with the Higgs field frozen at its vacuum configuration --- $i.e.$, $\Phi = \nu$. It is equivalent to replacing every Higgs field $\Phi$ in the action with its vacuum expectation value $\nu$. Since the $\R_{\xi}$ gauge fixing condition \eqref{Rxiguage} only depends on the vector gauge field $A_{\mu}$ and on the fluctuation of the scalar field $\varphi$ multiplied by the gauge parameter $\xi$, the Gribov procedure remains valid for the limit case $\xi \to 0$.

%In what follows, we will analyze the $SU(2)+$Higgs field in $d=3$ and $=4$ Euclidean space-time dimension. At first the Higgs field shall be considered in its fundamental representation, and in a second step its adjoint representation will be analysed. The next subsection is devoted to expose general $d-$dimensional results, and afterwards specific results of $d=3$ and $d=4$ will be presented.

%%%%%%%%%%%%%%%%%%%%%%%%%%%%%%%%%%%%%%%%%%%%%%%%%%%%%%%%%%%%%%%%%%%%
\section{The $SU(2)+$Higgs field}
\label{gfwithbroksym}
%%%%%%%%%%%%%%%%%%%%%%%%%%%%%%%%%%%%%%%%%%%%%%%%%%%%%%%%%%%%%%%%%%%%

\subsection{The Higgs field in the Fundamental representation: $d$-dimensional results}

In the present section nonperturbative effects of the $SU(2)+$Higgs model will be considered, by taking into account the existence of Gribov copies. The fundamental representation of the Higgs field, in $d=3$ and $d=4$, will be studied first. Subsequently, its adjoint representation, in $d=3$ and $d=4$, will be considered.
% Hereinafter, one should assume the Higgs field to be frozen at its vacuum configuration, as done in \cite{Fradkin:1978dv} --- in practice, the Higgs field self-coupling $\lambda$ will be made large enough in equation \eqref{SYMhiggs} --- and the Landau gauge limit, $\xi \to 0$, shall be taken at the very end of each computation.

Working in Euclidean spacetime, the starting action of the current model reads
\begin{equation}
S=\int \d^{d}x\left(\frac{1}{4}F_{\mu \nu }^{a}F_{\mu \nu }^{a}+
(D_{\mu }^{ij}\Phi ^{j})^{\dagger}( D_{\mu }^{ik}\Phi ^{k})+\frac{\lambda }{2}\left(
\Phi ^{\dagger}\Phi-\nu ^{2}\right) ^{2} - \frac{(\partial _{\mu }A_{\mu}^{a})^{2}}{2\xi} +\bar{c}^{a}\partial _{\mu }D_{\mu }^{ab}c^{b}\right)  \;, 
\label{SYMhiggs}
\end{equation}
where the covariant derivative is given by
\begin{equation}
D_{\mu }^{ij}\Phi^{j} =\partial _{\mu }\Phi^{i} -ig \frac{(\tau^a)^{ij}}{2}A_{\mu }^{a}\Phi^{j} \;,
\end{equation}
and the vacuum expectation value of the scalar field is $\left\langle \Phi^{i} \right\rangle ~=~ \nu\delta^{2i} $, with $i=1,2$,
%Introduced in the last section, $({\bar c}^a, c^a)$ are the anti-commuting Faddeev-Popov ghosts. The indices $i,j=1,2$ refer to the fundamental representation, and $\tau^a, a=1,2,3$, are the Pauli matrices.  The vacuum configuration, which minimizes the energy, is achieved by a constant scalar field parameterized as
%\begin{equation}
%\langle \Phi \rangle  = \left( \begin{array}{ccc}
%                                          0  \\
%                                          \nu
%                                          \end{array} \right)  \;,  \label{vevf}
%\end{equation}
so that all components of the gauge field acquire the same mass, $m^2= \frac{g^2\nu^2}{2}$.

By following the procedure described in the subsection \ref{Gribov's issue}, the restriction to the first Gribov region $\Omega$ relies on the computation of the ghost form factor, given by equation \eqref{nopole}, and on the enforcement of the no-pole condition, \eqref{ghstprop00}--\eqref{nopolestepfunct}. Since the presence of the scalar Higgs field does not influence the procedure of quantizing the gauge field, the non-local Gribov term, which is proportional to $\beta$, is not affected and one ends up with the action
\begin{eqnarray}
S\;' = S + \beta^* \sigma(0,A) -\beta^*   \;,
\label{effctactsu2}
\end{eqnarray}
with $S$ given by eq.\eqref{SYMhiggs}, the ghost form factor given by \eqref{nopole} and $\beta^{\ast}$ stands for the Gribov parameter that solves the gap equation.
%The no-pole condition is translated to the action by the insertion of a new parameter, called Gribov parameter, which is dynamically fixed by the gap equation, equation \eqref{gapeq}. The ghost form factor for the $SU(2)$ case becomes,
%\begin{eqnarray}
%\sigma(0,A) &=& \frac{1}{V} \frac{1}{d} \frac{2g^2}{3} \int\frac{ \d^d q}{(2 \pi)^d} A_\alpha^\ell(-q) A_\alpha^{\ell} (q) \frac{ 1 }{q^2}    \;.
%\label{nopole}
%\end{eqnarray}

\subsubsection{The $d$-dimensional gluon propagator and the gap equation}

In order to compute the gluon propagator up to one-loop order in perturbation theory let us follow the steps described in the subsection \ref{A sign of confinement from gluon propagator}. Therefore, we have
%\begin{eqnarray}
%S\,'_{quad} ~=~ \int \d^{d}x\left( \frac{1}{4} { \left(  \partial_\mu A^a_\nu -\partial_\nu A^a_\mu  \right)} ^2 - \frac{(\partial _{\mu }A_{\mu}^{a})^{2}}{2\xi}  + \frac{g^{2}\nu^{2}}{4}A_{\mu }^{a}A_{\mu }^{a}  \right) 
%\nonumber \\
% +  \frac{2g^2\beta}{3dV} \int\frac{ \d^d q}{(2 \pi)^4} A_{\mu}^{\ell}(-q) A_{\mu}^{\ell} (q) \frac{ 1 }{q^2}  - \beta  \;. 
%\label{quadf}
%\end{eqnarray}
%By changing to Fourier space and after a few algebraic manipulations, one gets for the partition function,
%\begin{equation}
%Z_{quad} ~=~ \int \frac{\d\beta  e^{\beta  }}{2\pi i\beta } [\d A] \; \exp \left\{  -\frac{1}{2} \int \frac{\d^{d}q}{(2\pi )^{d}}A_{\mu }^{a }(q)\mathcal{P}_{\mu \nu }^{ab }A_{\nu }^{b }(-q)   \right\}\;,  
%\label{Zq1f}
%\end{equation}
%with
%\begin{equation}
%\mathcal{P}_{\mu \nu }^{ab } =\delta ^{ab }\left[  \delta _{\mu \nu }\left( q^{2}+\frac{\nu^{2}g^{2}}{2}\right) +\left( \frac{1}{\xi } -1  \right) q_{\mu }q_{\nu } +  \frac{4g^2\beta}{3dV} \frac{1}{q^2} \delta _{\mu \nu }  \right]   \;.  
%\label{Pf}
%\end{equation}
%The gluon propagator may be taken immediately from the partition function \eqref{Zq1f} by constructing the generating functional. After all, the whole process boils down to finding the inverse of the operator \eqref{Pf} and taking $\xi \to 0$. At the end we must have,
\begin{equation}
\left\langle A_{\mu}^{a}(q)A_{\nu }^{b}(-q)\right\rangle ~=~  \delta^{ab}\frac{ q^2}{q^{4} + \frac{g^{2}\nu^{2}}{2} q^2
+  \frac{ 4g^2\beta^* }{3dV} }\left( \delta _{\mu\nu }-\frac{q_{\mu }q_{\nu }}{q^{2}}\right)   \;,
\label{propf0}
\end{equation}
whereby $\beta^{\ast}$ solves the gap equation, which is obtained by the means of the subsection \ref{Gribovimplementation}. After computing the Gaussian integral of the partition function and taking the trace over all indices, one ends up with the following partition function,
%At first one should integreat out the partition function \eqref{Zq1f} and, after that, take the trace over each index's space of the operator $\mathcal{P}_{\mu \nu }^{ab }$. After taking the trace one gets,
%The gap equation, needed to attach a dynamical meaning to the Gribov parameter $\beta$, may be derived from \eqref{Zq1f} by integrating out the gauge field --- which is simple since it is a Gaussian integral and, additionally, this process is quite the similar to that one developed in section \ref{Gribovimplementation} --- leading to
%\begin{equation}
%Z_{quad}  ~=~ \int {\frac{\d \beta}{2\pi i} }
%e^{({\beta} -\ln\beta)} \left(\det\mathcal{P}^{ab}_{\mu\nu}\right)^{-\frac{1}{2}} \;.
%\label{Zq2f}
%\end{equation}
%Replicating the process detailed in \eqref{Zq2f00} we could find
%\begin{equation}
%\left( \det \mathcal{P}_{\mu \nu }^{ab }\right) ^{-\frac{1}{2}} ~=~ \exp \left[ -\frac{3(d-1)V}{2}\int {\frac{\d^{d}q}{(2\pi )^{d}}\ln \left(
%q^{2}  +   \frac{g^{2}\nu^{2}}{2}  +  \frac{4g^2\beta}{3dV}   \frac{1}{q^{2}}\right) }\right] \;,
%\label{traceoperator}
%\end{equation}
%so that the partition function now reads,
%by taking the trace over all indices of the operator $\mathcal{P}^{ab}_{\mu\nu}$. Particularly, the factor $d-1$ comes from the trace over the space-time indices after taking the the Landau gauge $\xi \to 0$. Therefore, the partition function reads
\begin{equation}  \label{Zf}
Z_{quad} ~=~ \int \frac{\d\beta}{2\pi i}\; e^{f(\beta)} ~=~ \e^{-V{\cal E}_{v}}   \;,
\end{equation}
whereby one reads the free-energy,
\begin{equation}
f(\beta)  ~=~  \beta - \ln \beta - \frac{3(d-1)V}{2} \int \frac{\d^d k}{(2\pi)^d} \; \ln\left( k^2 + \frac{g^2\nu^2}{2} +  \frac{4g^2\beta}{3dV} \frac{1}{k^2} \right)  \;, \label{ff}
\end{equation}
which is equivalent to \eqref{minusfreenergy}. In the thermodynamic limit the integral of equation \eqref{Zf} may be solved through the saddle-point approximation \eqref{gapeq} leading to the gap equation\footnote{We remind here that the derivative of the term ${\ln\beta}$  in expression \eqref{ff} will be neglected, for the derivation of the gap equation, eq.\eqref{gapf}, when taking the thermodynamic limit.}
\begin{equation}
\frac{2(d-1)}{d}g^2 \int \frac{\d^d q}{(2\pi)^d} \frac{1}{ q^{4} + \frac{g^{2}\nu^{2}}{2} q^2  +   \frac{2g^2\beta^{\ast}}{3dV} }  = 1  \;.
\label{gapf}
\end{equation}

Let us now carry out the analysis of the gluon propagator for $d=3$ and $d=4$, after solving the gap equation.

\subsubsection{The $d=3$ case}

Now, let us proceed with the solution  of the gap equation \eqref{gapf}. Since we are working in $d=3$, the gap equation contain a finite integral, easy to be computed, leading to
\begin{equation}
 \frac{4g^2}{3dV} \beta^{\ast} = \frac{1}{4} \left( \frac{g^2\nu^2}{2} -  \frac{g^4}{9\pi^2} \right)^2 \;. \label{solf}
\end{equation}
As done in the previous section, the analysis of the gluon propagator could be simplified by making explicit use of its poles. Namely,
\begin{equation}
\left\langle A_{\mu }^{a}(q)A_{\nu }^{b}(-q)\right\rangle
=  \frac{\delta^{ab}}{m^2_+-m^2_-}\left(  \frac{m^2_+}{q^2+m^2_+} -\frac{m^2_{-}}{q^2+m^2_-}  \right)
\left( \delta _{\mu\nu }-\frac{q_{\mu }q_{\nu }}{q^{2}}\right)  \;,  \label{decf}
\end{equation}
with
\begin{equation}
m^2_+ = \frac{1}{2} \left( \frac{g^2\nu^2}{2}  + \sqrt{\frac{g^6}{9\pi^2} \left(\nu^2-\frac{g^2}{9\pi^2}\right)}\; \right) \;,  \qquad m^2_- = \frac{1}{2} \left( \frac{g^2\nu^2}{2}  - \sqrt{\frac{g^6}{9\pi^2} \left(\nu^2-\frac{g^2}{9\pi^2}\right)} \;\right)
\label{massesf} \;.
\end{equation}
%and
%\begin{equation}
%{\cal F}_{+} = \frac{m^2_+}{m^2_+-m^2_-}  \;, \qquad {\cal F}_{-} = \frac{m^2_-}{m^2_+-m^2_-}  \label{residf} \;.
%\end{equation}
In this way, we may distinguish two regions in the $(\nu^2,g^2)$ plane:
\begin{itemize}
\item[\it i)] when $g^2 < 9\pi^2 \nu^2$ both masses $(m^2_+,m^2_-)$ are positive, as well as the residues. The gluon propagator, eq.\eqref{decf}, decomposes into two Yukawa modes. However, due to the relative minus sign in expression \eqref{decf} only the heaviest mode with mass $m^2_+$ represents a physical mode. We see thus that, for $g^2 < 9\pi^2 \nu^2$, all components of the gauge field exhibit a physical massive mode with mass $m^2_+$. This region is what can be called a Higgs phase. 

Let us also notice that, for the particular value $g^2=\frac{9\pi^2}{2}\nu^2$, corresponding to a vanishing Gribov parameter $\beta=0$, the unphysical Yukawa mode in expression  \eqref{decf} disappears, as $m^2_-~=~0$. As a consequence, the gluon propagator reduces to that of a single physical mode with mass $\frac{9\pi^2}{4}\nu^4$.
%, namely
%\begin{equation}
%\left\langle A_{\mu }^{a}(q)A_{\nu }^{b}(-q)\right\rangle
%=\delta^{ab}   \left( \delta _{\mu\nu }-\frac{q_{\mu }q_{\nu }}{q^{2}}\right) \frac{1}{q^2 +\frac{9\pi^2}{4}\nu^4 } \;.  \label{decff}
%\end{equation}
\item[\it ii)] when $g^2> 9 \pi^2 \nu^2$, the masses $(m^2_+,m^2_-)$ become complex. In this region, the gluon propagator, eq.\eqref{decf}, becomes of the Gribov type, displaying complex conjugate poles. All components of the gauge field become thus unphysical. This region corresponds to the confining phase.
\end{itemize}
%Summarizing, when the Higgs field is in the fundamental representation, a Higgs phase is detected for $g^2 < 9\pi^2 \nu^2$. When $g^2> 9 \pi^2 \nu^2$, the confining phase emerges. Concerning the trustworthiness of the results, completely analogous comments as in the adjoint case apply here as well.

\subsubsection{The $d=4$ case}
With quite the same process as for the $d=3$ case, let us analyse the poles of the gauge field propagator by solving the gap equation for $d=4$. To that end the following decomposition becomes useful
\begin{equation}
q^4 +  \frac{g^{2}\nu^{2}}{2} q^2
+ \frac{g^2}{3} \beta   ~=~ (q^2+m^2_+) (q^2+m^2_-) \;,  
\label{dec1}
\end{equation}
with
\begin{equation}
m^2_+ ~=~ \frac{1}{2} \left(\frac{g^2 \nu^2}{2} + \sqrt{\frac{g^4\nu^4}{4}  -\frac{4g^2}{3} \beta^*} \;  \right) \;,  \qquad    m^2_- ~=~ \frac{1}{2} \left(\frac{g^2 \nu^2}{2} -\sqrt{\frac{g^4\nu^4}{4}  -\frac{4g^2}{3} \beta^*} \;  \right)  \;.
\label{roots1}
\end{equation}
Making use of the $\MSbar$ renormalization scheme in $d=4-\varepsilon$ 
%and of the standard integral
%\begin{equation}
%\int \frac{\d^d p}{(2\pi)^d} \frac{1}{p^2+\rho^2} ~=~ - \frac{\rho^2}{16\pi^2} \frac{2}{\bar \beta} + \frac{\rho^2}{16\pi^2} \left( \ln\frac{\rho^2}{{\omu}^2} - 1  \right) \;,
%\label{intd2f}
%\end{equation}
the gap equation \eqref{gapf} becomes
\begin{equation}
\left[1 + \frac{m^2_{-}}{m^2_+ -  m^2_-}\; \ln\left( \frac{m^2_-}{{\omu}^2} \right)  - \frac{m^2_+}{m^2_+ -  m^2_-}\; \ln\left( \frac{m^2_+}{{\omu}^2} \right)  \right] ~=~ \frac{32\pi^2}{3g^2}  \;. 
\label{gapf1}
\end{equation}
After a suitable manipulation we get a more concise expression for the gap equation
%In order to analyze this equation we rewrite it in a more suitable way, {\it i.e.}
%\begin{eqnarray}
% \frac{m^2_-}{m^2_+ -  m^2_-}\; \ln\left( \frac{m^2_-}{\omu^2} \right)  - \frac{m^2_+}{m^2_+ -  m^2_-}\; \ln\left( \frac{m^2_+}{\omu^2} \right)   
%&=& - \frac{m^2_+ - m^2_-}{m^2_+ -  m^2_-} \left( 1- \frac{32\pi^2}{3g^2} \right)  
%\nonumber \\
%&=&  \frac{m^2_+ - m^2_-}{m^2_+ -  m^2_-}\; \ln \left(  e^{- \left( 1- \frac{32\pi^2}{3g^2} \right)}  \right)   \;,  
%\label{gapf2}
%\end{eqnarray}
%so that
%\begin{equation}
% m^2_-\; \ln\left( \frac{m^2_-}{\omu^2  e^{\left( 1- \frac{32\pi^2}{3g^2} \right)} } \right) ~=~  m^2_+\; \ln\left( \frac{m^2_+}{\omu^2  e^{\left( 1- \frac{32\pi^2}{3g^2} \right)} } \right) \;,
%\label{gapf3}
%\end{equation}
%whose final form can be written as
\begin{equation}
2 \sqrt{1-\zeta}\; \ln(a) = -  \left( 1 +  \sqrt{1-\zeta} \right) \; \ln\left( 1 +  \sqrt{1-\zeta} \right) +  \left( 1 -  \sqrt{1-\zeta} \right) \; \ln\left( 1 - \sqrt{1-\zeta} \right)  \;, 
\label{gapd1}
\end{equation}
where we have introduced the dimensionless variables
\begin{equation}
a ~=~ \frac{g^2 \nu^2}{4 \omu^2 e^{\left( 1 -\frac{32\pi^2}{3g^2} \right)}} \;, \qquad  \qquad \zeta ~=~ \frac{16}{3} \frac{\beta^*}{g^2\nu^4}  \geq0 \;, \label{vb1}
\end{equation}
with $0 \le \zeta < 1$ in order to have two real, positive, distinct roots $(m^2_+, m^2_-)$.
% It is worth underlining that the renormalization scale $\omu$ could be exchanged in favour of the invariant scale  $\lms$, defined at one-loop as
%\begin{equation}
%\lms^2  =  \omu^2 \;e^{\frac{1}{\beta_0} \frac{1}{ g^2(\omu) } }  \label{Lambda} \;,
%\end{equation}
%with $\beta_0$ given by \cite{Gross:1973ju,Pickering:2001aq}
%\begin{equation}
%\beta  = - g^3 \beta_0 + O(g^5) \;, \qquad  \beta_0 = \frac{1}{16\pi^2} \left( \frac{11}{3} N - \frac{1}{6} T  \right)  \;,
%\end{equation}
%where $T$ is the Casimir of the representation of the Higgs field equaling $T=\frac{1}{2}$, resp.~$T=2$, for the fundamental, resp.~adjoint, representation of $SU(2)$.
For $\zeta >1$, the roots $(m^2_+, m^2_-)$ become complex conjugate, and the gap equation takes the form
\begin{equation}
2 \sqrt{\zeta -1} \; \ln(a) =   -2 \; \arctan\left({\sqrt{\zeta-1}}\; \right)   - \sqrt{\zeta-1} \; \ln\;\zeta  \;. \label{c1}
\end{equation}
Moreover, it is worth noticing that both expressions \eqref{gapd1},\eqref{c1} involve only one function, {\it i.e.} they can be written as
\begin{equation}
2 \; \ln(a) = g(\zeta)  \;, \label{g}
\end{equation}
where for $g(\zeta)$ we might take
\begin{equation}
g(\zeta) =    \frac{1}{ \sqrt{1-\zeta}} \left(
- \left( 1 +  \sqrt{1-\zeta} \right) \; \ln\left( 1 +  \sqrt{1-\zeta} \right) +  \left( 1 -  \sqrt{1-\zeta} \right) \; \ln\left( 1 - \sqrt{1-\zeta} \right)  \right) \;,
\label{gex}
\end{equation}
which is a real function of the variable $\zeta \ge 0$. Expression \eqref{c1} is easily obtained from \eqref{gapd1} by rewriting it in the region $\zeta>1$.  In particular, it turns out that the function $g(\zeta)\leq -2\ln 2$ for all $\zeta \ge 0$, and strictly decreasing. As consequence, for each value of $a<\frac{1}{2}$, equation \eqref{g} has always a unique solution with $\zeta>0$.  Moreover, it is easy to check that $g(1)=-2$. Therefore,  we can distinguish ultimately three regions, namely
\begin{itemize}
\item[(a)]  when $a>\frac{1}{2}$,  eq.\eqref{g}  has no solution for $\zeta$. Since the gap equation \eqref{gapf} has been  obtained by applying the saddle point approximation in the thermodynamic limit, we are forced to set $\beta^{\ast}=0$.  This means that, when $a>\frac{1}{2}$, the dynamics of the system is such that the restriction to the Gribov region cannot be consistently implemented.  As a consequence, the standard Higgs mechanism takes place, yielding three components of the gauge field with mass $m^{2} ~=~ \frac{g^{2}\nu^{2}}{2}$. Note that, for sufficiently weak coupling $g^2$, $a$ will unavoidably be larger than $\frac{1}{2}$.
%, according to
%\begin{equation}
%\left\langle A_{\mu }^{a}(q)A_{\nu }^{b}(-q)\right\rangle
%=\delta^{ab}\frac{1}{q^{2} + \frac{g^{2}\nu^{2}}{2}  }\left( \delta _{\mu
%\nu }-\frac{q_{\mu }q_{\nu }}{q^{2}}\right)  \label{propfY} \;.
%\end{equation}

\item[(b)] when $\frac{1}{e}<a<\frac{1}{2}$, equation  \eqref{g} has a solution for  $0 \le \zeta <1$. In this region, the roots $(m^2_+, m^2_-)$  are real and the gluon propagator decomposes into the sum of two terms of the Yukawa type:
\begin{equation}
\left\langle A_{\mu }^{a }(q)A_{\nu }^{b }(-q)\right\rangle
=\frac{\delta^{ab}}{m^2_+-m^2_-}  \left(   \frac{m^2_+}{q^2+m^2_+} -   \frac{m^2_-}{q^2+m^2_-}   \right) 
 \left( \delta _{\mu
\nu }-\frac{q_{\mu }q_{\nu }}{q^{2}}\right)  \label{ffin} \;.
\end{equation}
%where
%\begin{equation}
%{\cal F}_+ = \frac{m^2_+}{m^2_+-m^2_-}  \;, \qquad {\cal F}_- = \frac{m^2_-}{m^2_+-m^2_-} \;. \label{rf}
%\end{equation}
Moreover, due to the relative minus sign in eq.\eqref{ffin} only the component ${\cal F}_+$ represents a physical mode.
\item[(c)]  for $a<\frac{1}{e}$, equation \eqref{g} has a solution for  $\zeta>1$. This scenario will always be realized if $g^2$ gets sufficiently large, i.e.~at strong coupling. In this region the roots  $(m^2_+, m^2_-)$  become complex conjugate and the gauge boson propagator is of the Gribov type, displaying complex poles.  As usual, this can be interpreted as the confining region.
\end{itemize}
In summary, we clearly notice that at sufficiently weak coupling, the standard Higgs mechanism will definitely take place, as $a>\frac{1}{2}$, whereas for sufficiently strong coupling, we always end up in a confining phase because then $a<\frac{1}{2}$.

Having obtained these results, it is instructive to go back where we originally started. For a fundamental Higgs, all gauge bosons acquire a mass that screens the propagator in the infrared. This effect, combined with a sufficiently small coupling constant, will lead to a severely suppressed ghost self energy, i.e.~the average of \eqref{nopole} (to be understood after renormalization, of course). If the latter quantity will a priori not exceed the value of 1 under certain conditions --- {\it i.e.}, satisfying the no-pole condition --- the theory is already well inside the Gribov region and there is no need to implement the restriction. Actually, the failure of the Gribov restriction for $a>\frac{1}{2}$ is exactly because it is simply not possible to enforce that $\sigma(0)=1$. Perturbation theory in the Higgs sector is \emph{in se} already consistent with the restriction within the 1st Gribov horizon. Let us verify this explicitly by taking the average of \eqref{nopole} with, as tree level input propagator, a transverse Yukawa gauge field with mass $m^2=\frac{g^2\nu^2}{2}$. Using that there are 3 transverse directions\footnote{We have been a bit sloppy in this paper with the use of dimensional regularization. In principle, there are $3-\epsilon$ transverse polarizations in $d=4-\epsilon$ dimensions. Positive powers in $\epsilon$ can (and will) combine with the divergences in $\epsilon^{-1}$ to change the finite terms. However, as already pointed out before, a careful renormalization analysis of the Gribov restriction is possible, see e.g.~\cite{Dudal:2010fq,Vandersickel:2012tz} and this will also reveal that the ``1'' in the Gribov gap equation will receive finite renormalizations, compatible with the finite renormalization in e.g.~$\sigma(0)$, basically absorbable  in the definition of $a$. } in $4d$, we have
%\begin{eqnarray}
%\sigma(0) =\frac{3g^{2}}{2}\int \frac{d^{4}q}{(2\pi )^{4}}\frac{1}{q^2(q^2+\frac{g^2\nu^2}{2})}=-\frac{3}{\nu^2}\int\frac{d^4q}{(2\pi^4)}\frac{1}{q^2+\frac{g^2\nu^2}{2}}=-\frac{3g^2}{32\pi^2}\left(\ln\frac{g^2\nu^2}{2\omu^2}-1\right)\;.
%\end{eqnarray}
%Introducing $a$ as in \eqref{vb1}, we may reexpress the latter result as
\begin{eqnarray}
\sigma(0) =1-\frac{3g^2}{32\pi^2}\ln(2a)\;.
\end{eqnarray}
For $a>\frac{1}{2}$, the logarithm is positive and it is then evident that $\sigma(0)$ will not cross $1$, indicating that the theory already is well within the first Gribov horizon.

Another interesting remark is at place concerning the transition in terms of a varying value of $a$. If $a$ crosses $\frac{1}{e}$, the imaginary part of the complex conjugate roots becomes smoothly zero, leaving us with 2 coinciding real roots, which then split when $a$ grows. At $a=\frac{1}{2}$, one of the roots and its accompanying residue vanishes, to leave us with a single massive gauge boson. We thus observe that all these transitions are continuous, something which is in qualitative correspondence with the theoretical lattice predictions of the classic work \cite{Fradkin:1978dv} for a fundamental Higgs field that is ``frozen'' ($\lambda\to\infty$). Concerning the somewhat strange intermediate phase, {\it i.e.}  the one with a Yukawa propagator with a negative residue, eq.\eqref{ffin}, we can investigate in future work in more detail the asymptotic spectrum based on the BRST tools developed in \cite{Dudal:2012sb} when the local action formulation of the Gribov restriction is implemented. Recent works on the lattice confirm the existence of a cross-over region, where there is no line separating the ``phases'', as e.g. \cite{Maas:2013aia,Maas:2014pba} where the authors work in the non-aligned minimal Landau gauge and observe the transition between a QCD-like phase and a Higgs-like phase, in a region away from the cross-over region.

\subsubsection{The vacuum energy, phase transition and how trustworthy are the results?}
Let us look at the vacuum energy ${\cal E}_v$ of the system, which  can easily be read off from expression \eqref{Zf}, namely
\begin{equation}
{\cal E}_v = - \beta^* + \frac{9}{2} \int \frac{d^4k}{(2\pi)^4} \; \ln\left( k^2 + \frac{g^2\nu^2}{2} +\frac{\beta^*}{3} \frac{g^2}{k^2} \right)  \;, \label{ev}
\end{equation}
where $\beta^*$ is given by the gap equation \eqref{gapf}. Making use of the $\MSbar$ renormalization scheme, the vacuum energy may be written as:
%Making use of
%\begin{equation}
%\int \frac{d^dp}{(2\pi)^d} \; \ln(p^2 + m^2)  = - \frac{m^4}{32\pi^2} \left( \frac{2}{\bar \varepsilon}  -  \ln{\frac{m^2}{{\bar \mu}^2}} + \frac{3}{2} \right) \;,  \label{intdl}
%\end{equation}
%it is very easy to write down the vacuum energy:
\begin{itemize}
\item  for $a<\frac{1}{2}$, we have
\begin{eqnarray}
\frac{8}{9 g^4\nu^4}\; {\cal E}_v & = &  \frac{1}{32\pi^2} \left( 1 - \frac{32\pi^2}{3g^2} \right) - \frac{1}{2} \frac{\zeta}{32\pi^2} + \frac{1}{4}\frac{1}{32\pi^2} \left(  (4-2\zeta)\left( \ln( a) -\frac{3}{2} \right) \right)  \\ \nonumber & + &  \frac{1}{4}\frac{1}{32\pi^2} \left(   \left( 1+ \sqrt{1-\zeta} \right)^2 \ln  \left( 1+ \sqrt{1-\zeta} \right)
+ \left( 1- \sqrt{1-\zeta} \right)^2 \ln  \left( 1- \sqrt{1-\zeta} \right)
 \right)     \;, \label{v1}
\end{eqnarray}
where $\zeta$ is obtained through eqs.\eqref{g},\eqref{gex}. 
%Let us also give the expressions of the first two derivatives of ${\cal E}_v(a)$  with respect to $a$:
% From the gap equation \eqref{g},  we easily get
%\begin{equation}
%\frac{\partial \zeta}{\partial a} = \frac{2}{a} \frac{1}{g'(\zeta)}  \;. \label{saa}
%\end{equation}
%Therefore
%\begin{eqnarray}
%\frac{ \partial }{\partial a} \left[\frac{8}{9 g^4\nu^4}\; {\cal E}_v \right] & = & \frac{1}{64\pi^2} \frac{1}{a} (2-\zeta) \;, \nonumber \\
%\frac{ \partial^2 }{\partial a^2} \left[\frac{8}{9 g^4\nu^4}\; {\cal E}_v \right] & =& \frac{1}{64\pi^2} \frac{1}{a^2} \left(\zeta - 2 - \frac{2}{g'({\zeta})} \right)\;. \label{d12}
%\end{eqnarray}

\item for $a>\frac{1}{2}$,
\begin{equation}
\frac{8}{9 g^4\nu^4}\; {\cal E}_v  =   \frac{1}{32\pi^2} \left( 1 - \frac{32\pi^2}{3g^2} \right)  + \frac{1}{32\pi^2} \left(  \left( \ln( a) -\frac{3}{2} \right) \right)   +   \frac{1}{32\pi^2} \ln2    \;. \label{v2}
\end{equation}
\end{itemize}
%Owing to the fact that $g'(0) =- \infty$, it turns out that
From these expressions we could check that the vacuum energy ${\cal E}_v(a)$ is a continuous  function of the variable $a$, as well as its first and second derivative, and that the third derivative develops a jump at $a=\frac{1}{2}$. We might be tempted to interpret this is indicating a third order phase transition at $a=\frac{1}{2}$. The latter value actually corresponds to a line in the $(g^2,\nu)$ plane according to the functional relation \eqref{vb1}. However, we should be cautious to blindly interpret this value. It is important to take a closer look at the validity of our results in the light of the made assumptions. More precisely, we implemented the restriction to the horizon in a first order approximation, which can only be meaningful if the effective coupling constant is sufficiently small, while simultaneously emerging logarithms should be controlled as well. In the absence of propagating matter, the expansion parameter is provided by $y\equiv\frac{g^2N}{16\pi^2}$ as in pure gauge theory. The size of the logarithmic terms in the vacuum energy (that ultimately defines the gap equations) are set by $m_+^2\ln\frac{m_+^2}{\omu^2}$ and $m_-^2\ln\frac{m_-^2}{\omu^2}$. A good choice for the renormalization scale would thus be $\omu^2\sim |m_+^2|$: for (positive) real masses, a fortiori we have $m_-^2<m_+^2$ and the second log will not get excessively large either because $m_-^2$ gets small and the pre-factor is thus small, or $m_-^2$ is of the order of $m_+^2$ and the log itself small. For complex conjugate masses, the size of the log is set by the (equal) modulus of $m_\pm^2$ and thus both small by our choice of scale.

Let us now consider the trustworthiness, if any, of the $a=\frac{1}{2}$ phase transition point.  For $a\sim\frac{1}{2}$, we already know that $\zeta\sim 0$, so a perfect choice is $\omu^2\sim m_+^2\sim\frac{g^2\nu^2}{2}$. Doing so, the $a$-equation corresponds to
\begin{equation}\label{aeq}
    \frac{1}{2}\sim e^{-1+\frac{4}{3y}}
\end{equation}
so that $y\sim 4$. Evidently, this number is thus far too big to associate any meaning to the ``phase transition'' at $a=\frac{1}{2}$. Notice that there is no problem for the $a$ small and $a$ large region. If $\nu^2$ is sufficiently large and we set $\omu^2\sim \frac{g^2\nu^2}{2}$ we have a small $y$, leading to a large $a$, i.e.~the weak coupling limit without Gribov parameter and normal Higgs-like physics. The logs are also well-tempered.    For a small $\nu^2$, the choice $\omu^2\sim\sqrt{g^2\theta^*}$ will lead to
\begin{equation}\label{aeq2}
a\sim (\textrm{small number})e^{-1+\frac{4}{3y}}
\end{equation}
so that a small $a$  can now be compatible with a small $y$, leading to a Gribov parameter dominating the Higgs induced mass, the ``small number'' corresponds to $\frac{g^2\nu^2}{\sqrt{g^2\theta^*}}$. Due to the choice of $\omu^2$, the logs are again under control in this case.

Within the current approximation, we are thus forced to conclude that only for sufficiently small or large values of the parameter $a$ we can probe the theory in a controllable fashion. Nevertheless, this is sufficient to ensure the existence of a Higgs-like phase at large Higgs condensate, and a confinement-like region for small Higgs condensate. The intermediate $a$-region is more difficult to interpret due to the occurrence  of large logs and/or effective coupling. Notice that this also might make the emergence of this double Yukawa phase at $a=\frac{1}{e}\approx 0.37$ not well established at this point.

\subsection{The Higgs field in the ajoint representation: $d$-dimensional results}
\label{Adjrep}

Let us now assume the adjoint representation for the Higgs field in the action \eqref{SYMhiggs}.
%In this section we aim to use the action \eqref{SYMhiggs} but now for a different Higgs field representation. Namely, we are going to use the ajoint representation to the scalar Higgs field,
%\begin{equation}
%S=\int \d^{d}x\left(\frac{1}{4}F_{\mu \nu }^{a}F_{\mu \nu }^{a}+
%(D_{\mu }^{ij}\Phi ^{j})^{\dagger}( D_{\mu }^{ik}\Phi ^{k})+\frac{\lambda }{2}\left(
%\Phi ^{\dagger}\Phi-\nu ^{2}\right) ^{2} - \frac{(\partial _{\mu }A_{\mu}^{a})^{2}}{2\xi} +\bar{c}^{a}\partial _{\mu }D_{\mu }^{ab}c^{b}\right)  \;.
%\label{SYMhiggsadjoint}
%\end{equation}
In that case, the vacuum configuration which minimizes the energy is achieved by a constant scalar field satisfying
\begin{equation}
\left\langle \Phi ^{a}\right\rangle ~=~ \nu \delta ^{a3}   \;,
\label{higgsv}
\end{equation}
leading to the
%For the quadratic part of the action involving the gauge field $A^a_\mu$,  one easily obtains
%\begin{equation}
%S_{quad} ~=~ \int \d^{d}x   \;  \left( \frac{1}{4} { \left(  \partial_\mu A^a_\nu -\partial_\nu A^a_\mu  \right)} ^2 + b^a \partial_\mu A^a_\mu  +  \frac{g^{2}\nu ^{2}}{2}\left( A_{\mu }^{1}A_{\mu }^{1} + A_{\mu }^{2}A_{\mu}^{2}\right)  \right)  \;, 
%\label{quad}
%\end{equation}
%from which one could argue that the 
standard Higgs mechanism, so that the off-diagonal components of the gluon field $A^{\alpha}_\mu, \; \alpha=1,2$, acquires a mass $m_{H}^{2} ~=~ g^{2}\nu ^{2}$, {\it i.e.}
\begin{equation}
\label{gluonoff}
\left\langle A_{\mu }^{\alpha }(p)A_{\nu }^{\beta }(-p)\right\rangle =\frac{\delta ^{\alpha \beta }}{p^{2}+m_{H}^{2}}\left( \delta _{\mu \nu }-%
\frac{p_{\mu }p_{\nu }}{p^{2}}\right) \;,
\end{equation}
while the third component $A_\mu^3$ should remain massless,
\begin{equation}
\left\langle A_{\mu }^{3}(p)A_{\nu }^{3}(-p)\right\rangle =\frac{1}{p^{2}}%
\left( \delta _{\mu \nu }-\frac{p_{\mu }p_{\nu }}{p^{2}}\right) \;. 
\label{zm}
\end{equation}

However, as was pointed out by Polyakov  \cite{Polyakov:1976fu}, the theory exhibits a different behaviour. The action \eqref{SYMhiggs} admits classical solitonic solutions, known as the 't Hooft-Polyakov monopoles\footnote{ These configurations are instantons in Euclidean space-time.} which play a relevant role in the dynamics of the model. In fact, it turns out that these configurations give rise to a monopole condensation at weak coupling, leading to a confinement of the third component $A^3_\mu$, rather than to a Higgs type behaviour, eq.\eqref{zm}, a feature also confirmed by lattice numerical simulations  \cite{Nadkarni:1989na,Hart:1996ac}.

Since our aim is that of analysing the nonperturbative dynamics of the Georgi-Glashow model by taking into account the Gribov copies, let's follow the procedure described in the subsection \ref{Gribovimplementation}. Due to the presence of the Higgs field in the adjoint representation, causing a breaking of the global gauge symmetry, the ghost two-point function has to be decomposed into two sectors, diagonal and off-diagonal:
% Then, for the connected two-point ghost function $\mathcal{G}^{ab}(k;A)$ at first order in the gauge fields,  one finds  
%\begin{equation}
%\mathcal{G}^{ab}(k;A)  ~=~ \frac{1}{k^{2}}  \left( \delta ^{ab}-g^{2}\frac{k_{\mu}k_{\nu }}{k^{2}}  
%\int \frac{\d^{d}q}{(2\pi )^{d}} \;  \varepsilon^{amc}\varepsilon^{cnb}  \frac{1}{(k-q)^{2}}  \left( A_{\mu }^{m}(q)A_{\nu}^{n}(-q)\right) \right)  \label{Ghost} \;,
%\end{equation}
%where use has been made of the transversality condition $q_{\mu}A_{\mu}(q)=0$. 

%In order to correctly take into account the presence of the Higgs vacuum, eq.\eqref{higgsv}, we decompose  $\mathcal{G}^{ab}(k;A)$ into
\begin{equation}
\mathcal{G}^{ab}(k,A)=\left(
\begin{array}{cc}
\delta^{\alpha \beta}\mathcal{G}_{off}(k;A) & 0 \\
0 & \mathcal{G}_{diag}(k;A)
\end{array}
\right)
\end{equation}
where
\begin{eqnarray}
\mathcal{G}_{off}(k;A) 
%&=&  \frac{1}{k^{2}}   \left(   1+ \frac{g^{2}}{V}  \frac{k_{\mu }k_{\nu }}{2k^{2}}  
%\int \frac{\d^{d}q}{(2\pi )^{d}}  \;  \frac{1}{(q-k)^{2}}  \left( A_{\mu}^{\alpha }(q)A_{\nu }^{\alpha }(-q)+2A_{\mu }^{3}(q)A_{\nu }^{3}(-q)\right) \right)  
%\nonumber \\
&=&  \frac{1}{k^{2}}   \left( 1+\sigma _{off}(k;A)\right)       \approx \frac{1}{k^{2}}  \left( \frac{1}{1-\sigma
_{off}(k;A)}\right) 
\label{Goff} \;,  \\[5mm]
\mathcal{G}_{diag}(k;A) 
%&=&  \frac{1}{k^{2}}  \left( 1 +  \frac{g^{2}}{V} \frac{k_{\mu }k_{\nu}}{k^{2}}
%\int \frac{\d^{d}q}{(2\pi )^{d}} \;  \frac{1}{(q-k)^{2}}  \left( A_{\mu}^{\alpha }(q)A_{\nu }^{\alpha }(-q)\right) \right)  
%\nonumber \\
&=&  \frac{1}{k^{2}}  \left( 1+\sigma _{diag}(k;A)\right)  \approx \frac{1}{k^{2}}\left( \frac{1}{1-\sigma
_{diag}(k;A)}\right) \label{Gdiag} \;.
\end{eqnarray}

As we know, the quantities $\sigma_{off}(k;A), \; \sigma_{diag}(k;A)$ turn out to be  decreasing functions of the momentum $k$ and making use of the gauge field transversality, we have
\begin{eqnarray}
\sigma _{off}(0;A) &=&  \frac{g^{2}}{Vd}  \int \frac{\d^{d}q}{(2\pi )^{d}} \; \frac{\left( A_{\mu }^{3}(q)A_{\mu }^{3}(-q)+\frac{1}{2}A_{\mu }^{\alpha}(q)A_{\mu }^{\alpha }(-q)\right) }{q^{2}}  \;, 
\nonumber  \\
\sigma _{diag}(0;A) &=& \frac{g^{2}}{Vd}  \int {\frac{\d^{d}q}{(2\pi )^{d}} \; \frac{ \left( A_{\mu }^{\alpha }(q) A_{\mu }^{\alpha }(-q)\right) }{q^{2}}}    \;.
\label{sigma}
\end{eqnarray}
Therefore, the no-pole condition for the ghost function $\mathcal{G}^{ab}(k,A)$ is implemented by imposing that \cite{Gribov:1977wm,Vandersickel:2012tz,Sobreiro:2005ec}
\begin{eqnarray}
\sigma _{off}(0;A) &\leq &1\;, \nonumber \\
\sigma _{diag}(0;A) &\leq &1   \label{np} \;.
\end{eqnarray}

%By taking the limit $k \to 0$ of eqs.\eqref{Goff} and \eqref{Gdiag} and by making use of the property
%\begin{eqnarray}
%A_{\mu }^{a}(q)A_{\nu }^{a}(-q) &=&  \left( \delta _{\mu \nu }-\frac{q_{\mu }q_{\nu }}{q^{2}}\right) \omega (A)(q)   \nonumber \\
%&\Rightarrow &\omega (A)(q) ~=~ \frac{1}{d-1}A_{\lambda }^{a}(q)A_{\lambda }^{a}(-q)\;,
%\end{eqnarray}
%which follows from the transversality of the gauge field, $q_\mu A^a_\mu(q)=0$, we are able find the following expressions for $\sigma_{off}(0;A)$, and $\sigma_{diag}(0;A)$,

%Besides that, it is useful to remind that, for an arbitrary function $\mathcal{F}(p^2)$, we have
%\begin{equation}
%\int \frac{\d^{d}p}{(2\pi )^{d}}  \;  \left( \delta _{\mu \nu }-\frac{p_{\mu }p_{\nu }}{p^{2}} \right) 
%\mathcal{F}(p^2)  ~=~  \mathcal{A}  \;  \delta _{\mu \nu }  \label{a}
%\end{equation}
%where, upon contracting both sides with $\delta_{\mu\nu}$,
%\begin{equation}
%\mathcal{A}=\frac{d-1}{d}\int \frac{\d^{d}p}{(2\pi )^{d}}\mathcal{F}(p^2).
%\end{equation}

After that we need two different parameters to implement the no-pole condition in the action, leading us to the following action accounting for the Gribov ambiguities,
\begin{eqnarray}
S\;' = S + \beta^* \left( \sigma_{off}(0,A) - 1 \right) + \omega^* \left( \sigma_{diag}(0,A) - 1 \right)   \;.
\label{effctactadjoit}
\end{eqnarray}
In the action \eqref{effctactadjoit} $\beta^*$ and $\omega^*$ are given dynamically through its own gap equation.

\subsubsection{The $d$-dimensional gluon propagator and the gap equation}

The partition function accounting only for quadratic terms of the action \eqref{effctactadjoit} can be written as
%Just as was made before, the Gribov gap equation would be obtained by integrating out the equation
\begin{equation}
Z_{quad} ~=~ \int \frac{\d\beta e^{\beta }}{2\pi i\beta }\frac{\d\omega e^{\omega }} {2\pi i\omega }[\d A^{\alpha }][\d A^{3}] \; e^{-\frac{1}{2}  \int \frac{\d^{d}q}{(2\pi )^{d}}  \;  A_{\mu }^{\alpha }(q)\mathcal{P}_{\mu \nu }^{\alpha
\beta }A_{\nu }^{\beta }(-q)-\frac{1}{2}  \int \frac{\d^{d}q}{(2\pi )^{d}} \;  A_{\mu }^{3}(q)\mathcal{Q}_{\mu \nu }A_{\nu }^{3}(-q)},  
\label{Zq1}
\end{equation}
with
\begin{eqnarray}
\mathcal{P}_{\mu \nu }^{\alpha \beta } &=&  \delta^{\alpha \beta } \left(\delta _{\mu \nu } \left( q^{2}+\nu^{2}g^{2}\right) +\left( \frac{1}{\xi } -1 \right) q_{\mu }q_{\nu } +  \frac{2g^{2}}{Vd}\left( \beta +\frac{\omega }{2} \right) \frac{1}{q^{2}}\delta _{\mu \nu }\right)  \label{P}  \;, \nonumber \\
\mathcal{Q}_{\mu \nu } &=& \delta_{\mu \nu }\left( q^{2} - \frac{2\omega g^{2}}{Vd}\frac{1}{q^{2}}\right) +\left( \frac{1}{\xi }-1\right) q_{\mu }q_{\nu } \;,
\label{Q}
\end{eqnarray}
The parameter $\xi$ stands for the usual gauge fixing parameter, to be put to zero at the end in order to recover the Landau gauge. Evaluating  the inverse of the expressions \eqref{Q} and taking the limit $\xi\rightarrow 0$, the gluon propagators become
\begin{eqnarray}
\left\langle A_{\mu }^{3}(q)A_{\nu }^{3}(-q)\right\rangle &=&\frac{q^{2}}{q^{4}+  \frac{2\omega g^{2}}{Vd} }  \left( \delta _{\mu \nu }-\frac{q_{\mu }q_{\nu}}{q^{2}}\right)  \label{Pdiag} \;, 
\\
\left\langle A_{\mu}^{\alpha}(q) A_{\nu}^{\beta }(-q)\right\rangle  &=&  \delta^{\alpha\beta} 
\frac{q^{2}}{q^{2} \left( q^{2}+g^{2}\nu^{2}\right) +  \frac{2g^{2}}{Vd} \left(  \beta + \frac{\omega}{2}\right)}  
\left(\delta_{\mu\nu} - \frac{q_{\mu}q_{\nu}}{q^{2}}\right)  \;.
\label{NPoff} 
\end{eqnarray}
The off-diagonal sector of the gluon propagator can be put in a more convenient form by making explicit its poles. Namely,
\begin{eqnarray}
\left\langle A_{\mu}^{\alpha}(q) A_{\nu}^{\beta }(-q)\right\rangle 
% &=&  \delta^{\alpha\beta} 
%\frac{q^{2}}{q^{2} \left( q^{2}+g^{2}\nu^{2}\right) +  \frac{2g^{2}}{Vd} \left(  \beta + \frac{\omega}{2}\right)}  
%\left(\delta_{\mu\nu} - \frac{q_{\mu}q_{\nu}}{q^{2}}\right)
%\nonumber  \\
&=&  \frac{\delta ^{\alpha \beta }}{m^2_+-m^2_-}   \left(  \frac{m^2_+}{q^2+m^2_+} -\frac{m^2_{-}}{q^2+m^2_-}  \right)
\left( \delta _{\mu\nu }-\frac{q_{\mu }q_{\nu }}{q^{2}}\right)  \;,  
\label{NPoff_f1}
\end{eqnarray}
with
\begin{equation}
m^2_+ = \frac{g^2\nu^2 + \sqrt{g^4 \nu^4 - 4 \tau}}{2}  \;,  \qquad m^2_- = \frac{g^2\nu^2 - \sqrt{g^4 \nu^4 - 4 \tau}}{2} 
\;, \qquad  \tau = \frac{2g^2}{Vd} \left( \beta +\frac{\omega}{2} \right)  \;.\label{masses}
\end{equation}

Since the Gribov parameters $(\beta, \omega)$ are fixed dynamically through the gap equation, now we should integreat out the gauge field from equation \eqref{Zq1} and make use of the saddle-point approximation, in the thermodynamic limit, which will gives us two gap equations, enabling us to express $\beta$ and $\omega$ in terms of the parameters of the starting model, {\it i.e.} the gauge coupling constant $g$ and the {\it vev} of the Higgs field $\nu$. Performing the Gaussian integral of the partition function \eqref{Zq1} we get\footnote{In order to obtain eq. \eqref{Zq4} we had to take the functional trace of $\ln \mathcal{P}$ and $\ln \mathcal{Q}$ following the process described in subsection \ref{Gribovimplementation}. For more details take a look at \cite{Vandersickel:2012tz}.}
%\begin{equation}
%Z_{quad}=\int{\frac{d\beta}{2\pi i\beta}\frac{d\omega}{2\pi i\omega}}%
%e^{\beta}e^{\omega}\left(\det\mathcal{Q}_{\mu\nu}\right)^{-\frac{1}{2}%
%}\left(\det\mathcal{P}^{\alpha\beta}_{\mu\nu}\right)^{-\frac{1}{2}} \;,
%\label{Zq2}
%\end{equation}
%which, by making use of the process to exponentiate the functional determinant of \eqref{Zq2}, referred in \eqref{functdeterminant}, one gets
%Making use of
%\begin{equation}
%\left(\det \mathcal{A}_{\mu\nu}^{ab}\right)^{-\frac{1}{2}} ~=~ e^{-\frac{1}{2}%
%\ln \det \mathcal{A}_{\mu\nu}^{ab}} ~=~ e^{-\frac{1}{2}Tr \ln \mathcal{A}%
%_{\mu\nu}^{ab}} \;,
%\end{equation}
%for the determinants in expression (\ref{Zq2}) we get
%\begin{eqnarray}
%\left( \det \mathcal{Q}_{\mu \nu }\right) ^{-\frac{1}{2}} &=& \exp \left[
%-\frac{(d-1)V}{2} \int {\frac{\d^{d}q}{(2\pi )^{d}} \; \ln \left( q^{2} +  \frac{2\omega g^{2}}{Vd}  \frac{1}{q^{2}}\right) }\right] \;, \nonumber \\
%\left( \det \mathcal{P}_{\mu \nu }^{\alpha \beta }\right) ^{-\frac{1}{2}}
%&=&  \exp \left[ - \frac{2(d-1)V}{2}  \int {\frac{\d^{d}q}{(2\pi )^{d}} \; \ln \left(
%(q^{2}+g^{2}\nu^{2})  +  \frac{2g^{2}}{Vd}  \left( \beta  +  \frac{\omega }{2}\right)
%\frac{1}{q^{2}}\right) }\right] \;.
%\end{eqnarray}
%Then, the partition function \eqref{Zq2} may be rewritten as 
\begin{equation}  \label{Zq3}
Z_{quad}=\int{\frac{\d\beta}{2\pi i}\frac{\d\omega}{2\pi i}}e^{f(\omega,\beta)} \;,
\end{equation}
with
\begin{eqnarray}\label{Zq4}
f(\omega ,\beta ) &=& \beta +\omega -\ln \beta -\ln \omega - \frac{(d-1)V}{2}  \int {\frac{\d^{d}q}{(2\pi )^{d}}  \; \ln \left( q^{2}+\frac{2\omega g^{2}}{Vd}\frac{1}{q^{2}}\right)
} \nonumber  \\
&-&  \frac{2(d-1)V}{2}  \int {\frac{\d^{d}q}{(2\pi)^{d}}  \; \ln \left((q^{2}+g^{2}\nu^{2}) + \frac{g^{2}}{Vd} \left( 2\beta + \omega \right)
\frac{1}{q^{2}}\right) } \;.
\end{eqnarray}
Since in the thermodynamic limit, as mentioned in the section \ref{Gribovimplementation}, the integral \eqref{Zq3} can be solved through the saddle point approximation,
%\begin{equation}
%Z_{quad}\approx e^{f(\beta^*,\omega^*)} \;,
%\end{equation}
%with $\beta^*$ and $\omega^*$ being determined by the stationary conditions
\begin{equation}
\frac{\partial f}{\partial \beta^*}=\frac{\partial f}{\partial \omega^*}=0 \;,
\end{equation}
one gets the following two gap equations
%\footnote{We remind here that the terms $\log\beta$ and $\log \omega$ can be neglected in the derivation of the gap equations, eqs.\eqref{gap1} \eqref{gap2}, when taking the thermodynamic limit \cite{Gribov:1977wm,Sobreiro:2005ec,Vandersickel:2012tz}.}:
\begin{eqnarray}
\frac{4(d-1)g^{2}}{2d} \int \frac{\d^{d}q}{(2\pi )^{d}} \frac{1}{q^{4}+\frac{2\omega^{\ast}g^{2}}{d}}  &=&1 \;, 
\label{gap1} \\
\frac{4(d-1)g^{2}}{2d}  \int \frac{\d^{d}q}{(2\pi )^{d}}\left( \frac{1}{q^{2}(q^{2}+g^{2}\nu^{2})+g^{2}\left( \frac{2\beta ^{\ast}}{d}+\frac{\omega^{\ast}}{d}\right) }\right) &=&1 \;.
\label{gap2}
\end{eqnarray}
Therefore, $\beta^*$ and $\omega^*$ can be expressed in terms of the parameters $\nu,g$. To solve the gap equations the denominator of eq.\eqref{gap2} can be decomposed into its poles, which is similar to \eqref{NPoff}--\eqref{masses}.

%\subsubsection{Gluon propagators}

%Just as already done before, let's decompose the gluon propagator in order to simplify the analysis of its poles. In particular, the propagator \eqref{Pdiag} does not need to be decomposed, since it is quite similar to Gribov's propagator \eqref{Gribovprop1}. By the other side, the analysis of the off-diagonal sector propagator \eqref{NPoff} becomes easier if it is written as

%and
%\begin{equation}
%{\cal R}_{+} =   \;, \qquad {\cal R}_{-} = \frac{m^2_-}{m^2_+-m^2_-}  \label{residues} \;.
%\end{equation}

\subsection{The $d=3$ case}
\label{3dsu2}

%The first integral is easy to compute, giving $\omega^* $ as a function of $g$
%only

%In order to solve the second gap equation \eqref{gap2}  we compute the roots of the
%denominator:
%\begin{equation}
%m_{\pm }^{2}=\frac{-g^{2}\nu^{2}\pm \sqrt{g^{4}\nu^{4}-4\tau }}{2}  \label{p+-}  \;.
%\end{equation}%
%Notice that the roots are real when
%\begin{equation}
% \tau \leq \frac{g^{4}\nu ^{4}}{4}  \qquad \text{with} \qquad \tau =\beta^* \frac{2g^{2}}{3}+\omega^* \frac{g^{2}}{3} \;.
%\label{cond}
%\end{equation}
%After decomposition in partial fractions,  equation \eqref{gap2}  becomes
%\begin{eqnarray}
%\frac{4g^{2}\pi }{(2\pi )^{3}(q_{+}^{2}-q_{-}^{2})}\int_{0}^{\infty
%}dq\left( \frac{q^{2}}{(q^{2}-q_{+}^{2})}-\frac{q^{2}}{(q^{2}-q_{-}^{2})}%
%\right) &=&1 \;.
%\end{eqnarray}%
%Using the principal value prescription, this yields the final (finite) result
%\begin{equation}
%\frac{ig^{2}}{(4\pi )}\frac{1}{q_{+}-q_{-}}=1 \;.  \label{one}
%\end{equation}
%Making use of expression \eqref{p+-}, equation \eqref{one} gives $\tau$ as function of the parameters $(\nu,g)$, {\it i.e.}

%\subsubsection{Analysis of the gluon propagator}

In the three-dimensional case both gap equations cause not many difficulties to be solved, as there are no divergences to be treated. Namely, the first gap equation, eq.\eqref{gap1}, leads to the following result,
\begin{eqnarray}
\omega^*(g)= \frac{3}{2^{11}\pi^4} \;g^6        \;,  \label{omega}
\end{eqnarray}
while the second one, given by eq.\eqref{gap2}, leads to
\begin{equation}
\tau =\beta^* \frac{2g^{2}}{3}+\omega^* \frac{g^{2}}{3}  = \left[ \frac{1}{2}g^{2}\nu^{2}-\frac{g^{4}}{32\pi^2 }\right] ^{2}  \;. \label{feq}
\end{equation}

Now we can look at the gluon propagators, \eqref{Pdiag} and \eqref{NPoff}, and analyse the different regions in the $(g,\nu)$ plane. Let us start by the diagonal component $A^3_\mu$. Namely, we have
\begin{equation}
\left\langle A_{\mu }^{3}(q)A_{\nu }^{3}(-q)\right\rangle =\frac{q^{2}}{%
q^{4}+\frac{2\omega^* g^{2}}{3}}\left( \delta _{\mu \nu }-\frac{q_{\mu }q_{\nu
}}{q^{2}}\right)  \label{Pdiagf} \;.
% \qquad \omega^*(g)= \frac{3}{2^{11}\pi^4} \;g^6  \;.
\end{equation}
One observes that expression \eqref{Pdiagf} turns out to be independent from the $vev$ $\nu$ of the Higgs field, while displaying two complex conjugate poles. This gauge component is thus of the Gribov type. In other words, the mode $A^3_\mu$ is always confined, for all values of the parameters $g, \nu$. Concerning now the off-diagonal gluon propagator \eqref{NPoff}, after decomposing it into two Yukawa modes  \eqref{NPoff_f1}, we could find the following regions in the $(g,\nu)$ plane:

\begin{itemize}
\item[\it i)] 
when $g^2 < 32 \pi^2 \nu^2$, corresponding to $\tau < \frac{g^2\nu^2}{4}$, both masses $m^2_+, m^2_-$ are real, positive and  different from each other. Moreover, due to the presence of the relative minus sign in expression \eqref{NPoff_f1}, only the heaviest mode with mass $m^2_+$ represents a physical excitation --- {\it i.e.}, despite the existence of two real positive poles, $m^2_{+}$ and $m^{2}_{-}$, only the contribution related to the $m^{2}_{+}$ pole has physical meaning.

It is also worth observing that, for the particular value $g=16 \pi^2 \nu^2$, corresponding to $\tau=0$, the unphysical mode in the decomposition \eqref{NPoff_f1} disappears. Thus, for that particular value of the gauge coupling, the off-diagonal propagator reduces to a single physical Yukawa mode with mass $16\pi^2\nu^4$.
%, {\it i.e.}
%\begin{equation}
%\left\langle A_{\mu }^{\alpha }(q)A_{\nu }^{\beta }(-q)\right\rangle
%= \delta ^{\alpha \beta }  \left(  \frac{1}{q^2+16\pi^2\nu^4} \right)
%\left( \delta _{\mu\nu }-\frac{q_{\mu }q_{\nu }}{q^{2}}\right)  \;,  \label{off-Yuk}
%\end{equation}
\item[\it ii)] when $g^2>32\pi^2\nu^2$, corresponding to $\tau> \frac{g^2\nu^2}{4}$, all masses become complex and the off-diagonal propagator becomes of the Gribov type with two complex conjugate poles. This region, called Gribov region since all modes are of Gribov type, corresponds to a phase in which all gauge modes are said to be confined.
\end{itemize}
In summary, when the Higgs field is in the adjoint representation we could find two distinct regions. For $g^2<32\pi^2\nu^2$ the $A_3$ mode is confined while the off-diagonal propagator displays a physical Yukawa mode with mass $m^2_+$. This phase is referred to as the $U(1)$ symmetric phase \cite{Nadkarni:1989na,Hart:1996ac}. When $g^2>32\pi^2\nu^2$ all propagators are of the Gribov type, displaying complex conjugate poles leading to a confinement interpretation. According to \cite{Nadkarni:1989na,Hart:1996ac} this regime is referred to as the $SU(2)$ confined phase. 

Since our results were obtained in a semi-classical approximation ({\it i.e.}, lowest order in the loop expansion), let us comment on the validity of such approximation. In general, the perturbation theory is reliable when the \emph{effective coupling constant} is sufficiently small. The effective coupling depends, in $3d$, on the factor $\frac{g^2}{(4\pi)^{3/2}}$. However, since $g^2$ has mass dimension $1$ the effective coupling is not complete yet. In the presence of a mass scale $M$, the perturbative series --- for e.g.~the gap equation --- will organize itself automatically in a series in $G^2/M$. Let us analyse, for example, the case where $g^2 < 32 \pi^2 \nu^2$, the called the ``Higgs phase''. In this case the effective coupling will be sufficiently small when $\frac{g^2}{\nu^2(4\pi)^{3/2}}$ is small compared\footnote{The Higgs mass $\nu^2$ is then the only mass scale entering the game.} to $1$. Such condition is not at odds with the retrieved condition $g^2 < 32 \pi^2 \nu^2$. Next, assuming the coupling $g^2$ to get large compared to $\nu^2$, thereby entering the confinement phase with $cc$ masses, $g^2$ dominates everything, leading to a Gribov mass scale $\tau\propto g^8$, and an appropriate power of the latter will secure a small effective expansion parameter consistent with the condition $g^2 > 32 \pi^2 \nu^2$. We thus find that at sufficiently small and large values of $\frac{g^2}{\nu^2}$ our approximation and results are trustworthy.

\subsection{The $d=4$ case}

Let us start by considering the second gap equation, eq.\eqref{gap2}. Performing the decomposition described in eq.\eqref{masses} the referred gap equation becomes of the same form as the one obtained in the fundamental $d=4$ case, eq.\eqref{gapf1}. The difference between the fundamental and adjoint $d=4$ cases appears in the definition of the mass parameter $m^{2}_{\pm}$ (see eq.\eqref{roots1} and eq.\eqref{masses}, respectively). Namely,
%\begin{equation}
%q^4 + g^2 \nu^2 q^2 + \tau = (q^2+m^2_+) (q^2+m^2_-) \;, \qquad \tau = g^2 \left( \frac{\beta^*}{2}+\frac{\omega^*}{4} \right)  \;, \label{dec}
%\end{equation}
%with $m^{2}_{\pm}$ given by equation \eqref{masses}. Let us discuss first the case of two real,  positive, different roots, namely $0 \le \tau <  \frac{g^4\nu^4}{4}$. From
%\eqref{intd2f}, we reexpress the gap equation \eqref{gap2} as
\begin{equation}
\left(1 + \frac{m^2_-}{m^2_+ -  m^2_-}\; \ln\left( \frac{m^2_-}{{\omu}^2} \right)  - \frac{m^2_+}{m^2_+ -  m^2_-}\; \ln\left( \frac{m^2_+}{{\omu}^2} \right)  \right) = \frac{32\pi^2}{3g^2}  \;. \label{gap2bb}
\end{equation}
Introducing now the dimensionless variables\footnote{We introduced the renormalization group invariant scale $\lms$. }
\begin{eqnarray}
b ~=~  \frac{g^2\nu^2}{2 {\bar \mu}^2\; e^{\left(1-\frac{32\pi^2}{3g^2}\right)} } =\frac{1}{2\; e^{\left( 1 -\frac{272 \pi^2}{21 g^2} \right)}}  \;\frac{g^2\nu^2}{\lms^2  } \;, \qquad \text{and} \qquad 
 \xi  ~=~  \frac{4\tau}{g^4\nu^4}  \geq 0\;,
\label{vb}
\end{eqnarray}
with $0 \le \xi < 1 \;$. Proceeding as in the fundamental $d=4$ case, eq. \eqref{gap2bb} can be recast in the following form
\begin{equation}
2 \sqrt{1-\xi}\; \ln(b) = -  \left( 1 +  \sqrt{1-\xi} \right) \; \ln\left( 1 +  \sqrt{1-\xi} \right) +  \left( 1 -  \sqrt{1-\xi} \right) \; \ln\left( 1 - \sqrt{1-\xi} \right)  \;. \label{gapdd}
\end{equation}
or compactly,
\begin{equation}
2\ln b=g(\xi)\;. \label{gapddbis}
\end{equation}
Also here, in the adjoint $d=4$ case, eq.\eqref{gapddbis} remains valid also for complex conjugate roots, viz.~$\xi>1$. We are then led to the following cases.

\subsubsection{When $b<\frac{1}{2}$}
Using the properties of $g(\xi)$, it turns out that eq.\eqref{gapdd} admits a unique solution for $\xi$, which can be explicitly constructed with a numerical approach. More precisely, when the mass scale $g^2\nu^2$ is sufficiently smaller  than $\lms^2$, {\it i.e.}
\begin{equation}
g^2 \nu^2 < 2 \; e^{\left( 1 -\frac{272 \pi^2}{21 g^2} \right)} \; \lms^2   \;, \label{mh}
\end{equation}
we have what can be called the $U(1)$ confined phase. In fact, in this regime the gap equation \eqref{gap1} leads to a non-null $\omega^*$, so that the diagonal component of the gauge field is said to be of the Gribov type, {\it i.e.} with confinement interpretation.
%and \eqref{gap2} can be rewritten as
%\begin{eqnarray}
%3 \left( \frac{g^{2}}{2}\right) \int \frac{d^{4}q}{(2\pi )^{4}}\left( \frac{1}{%
%q^{4}+g^2 \frac{\omega^*}{2}}\right) &=&1\;, \label{ggapp1} \\
%3 \left( \frac{g^{2}}{2}\right) \int \frac{d^{4}q}{(2\pi )^{4}}\left( \frac{1%
%}{q^{2}(q^{2}+g^{2}\nu^{2})+g^2 (\frac{\beta^*}{2}+\frac{\omega^*}{4})}
%\right) &=&1\label{ggapp2} \;.
%\end{eqnarray}
%Moreover, making use of
%\begin{equation}
%\int \frac{d^dp}{(2\pi)^d} \frac{1}{p^4 + m^4}  = \frac{2}{\bar \varepsilon} \frac{1}{16\pi^2} - \frac{1}{16\pi^2} \left( \ln\frac{m^2}{\omu^2} - 1 \right) \;,  \label{intd1}
%\end{equation}
%the gap equation \eqref{ggapp1} gives
%\begin{equation}
%{\left( \frac{g^2 \omega^*}{2} \right)}^{1/2} = {\bar \mu}^2  e^{ \left(1 - \frac{32\pi^2}{3g^2} \right)}  = \lms^2\; e^{\left( 1 -\frac{272 \pi^2}{21 g^2} \right)}  \;.
%\end{equation}
%Therefore, for $b<\frac{1}{2}$, the $A^3_\mu$ component of the gauge field gets confined, exhibiting a Gribov-type propagator with complex poles, namely
%\begin{equation}
%\left\langle A_{\mu }^{3}(q)A_{\nu }^{3}(-q)\right\rangle =\frac{q^{2}}{%
%q^{4}+\frac{\omega^* g^{2}}{2}}\left( \delta _{\mu \nu }-\frac{q_{\mu }q_{\nu
%}}{q^{2}}\right)  \;. \label{Pdiag_conf}
%\end{equation}

On the other side, the second gap equation \eqref{gapdd} splits this region in the two subregions:
\begin{itemize}
\item[(i)]  when $\frac{1}{e}<b<\frac{1}{2}$ equation  \eqref{gapdd} has a single solution with $0 \le \xi <1$. In this region, the roots $(m^2_+, m^2_-)$  are thus real and  the off-diagonal propagator decomposes into the sum of two Yukawa propagators.
%\begin{equation}
%\left\langle A_{\mu }^{\alpha }(q)A_{\nu }^{\beta }(-q)\right\rangle
%=\delta ^{\alpha \beta } \left(  \frac{{\cal R}_+}{m^2+m^2_+} -    \frac{{\cal R}_-}{m^2+m^2_-}   \right)
%\left( \delta _{\mu
%\nu }-\frac{q_{\mu }q_{\nu }}{q^{2}}\right)  \label{NPoff_fin} \;,
%\end{equation}
%where
%\begin{equation}
%{\cal R}_+ = \frac{m^2_+}{m^2_+-m^2_-}  \;, \qquad {\cal R}_- = \frac{m^2_-}{m^2_+ - m^2_-} \;. \label{r}
%\end{equation}

However, due to the relative minus sign in eq.\eqref{NPoff_f1},  only the component with $m^{2}_+ $ pole can be associated to a physical mode, analogously as in the fundamental case. Due to the confinement of the third component $A^3_\mu$, this phase is recognized as the $U(1)$ confining phase. It is worth observing that it is also present in the $3d$ case, with terminology coined in \cite{Nadkarni:1989na}, see also \cite{Capri:2012cr}.

\item[(ii)]  for $b<\frac{1}{e}$, equation \eqref{gapdd} has a solution for  $\xi>1$. In this region the roots  $(m^2_+, m^2_-)$  become complex conjugate and the off-diagonal gluon propagator is of the Gribov type, displaying complex poles.  In this region all gauge fields display a propagator of the Gribov type. This is recognized as the $SU(2)$ confining region.
\end{itemize}
Similarly, the above regions are continuously connected when $b$ varies. In particular, for $b \stackrel{<}{\to} \frac{1}{2}$, we obtain $\xi=0$ as solution.

\subsubsection{The case $b>\frac{1}{2}$}
Let us consider now the case in which $b>\frac{1}{2}$. Here, there is no solution of the equation \eqref{gapdd} for the parameter $\xi$, as it follows by observing that the left hand side of eq.\eqref{gapdd} is always positive, while the right hand side is always negative. This has a deep physical consequence. It means that for a Higgs mass $m_{Higgs}^2 = g^2\nu^2$ sufficiently larger than $\lms^2$, {\it i.e.}
\begin{equation}
g^2 \nu^2 > 2 \; e^{\left( 1 -\frac{272 \pi^2}{21 g^2} \right)} \; \lms^2   \;, \label{mh}
\end{equation}
the gap equation \eqref{gap2} is inconsistent. It is then important to realize that this is actually the gap equation obtained by acting with $\frac{\p}{\p \beta}$ on the vacuum energy  ${\cal E}_v= -  f(\omega ,\beta )$.
So, we are forced to set $\beta^*=0$, and confront the remaining $\omega$-equation, viz.~eq.\eqref{gap1},
%:
%\begin{equation}
%\frac{3}{2}\left( \frac{g^{2}}{2}\right) \int \frac{d^{4}q}{(2\pi )^{4}}\left( \frac{1}{q^{4}+\frac{%
%\omega ^{\ast }g^{2}}{2}}+\frac{1}{q^{2}(q^{2}+g^{2}\nu^{2})+\frac{g^2\omega ^{\ast }}{4} }\right) =1 \;, \label{gap1n}
%\end{equation}
which can be transformed into
\begin{eqnarray}
4 \; \ln(b) & = &  \frac{1}{\sqrt{1-\xi}} \left[ -\left( 1 +  \sqrt{1-\xi} \right) \; \ln\left( 1 +  \sqrt{1-\xi} \right) +  \left( 1 -  \sqrt{1-\xi} \right) \; \ln\left( 1 - \sqrt{1-\xi} \right) \right.  \nonumber \\
 &\; \;&-\left. \sqrt{1-\xi}\; \ln\xi - \sqrt{1-\xi} \ln 2\right]\equiv h(\xi)
 \label{gapddn}
\end{eqnarray}
after a little algebra, where $\xi=\frac{\omega^*}{g^4\nu^4}$. The behaviour of $h(\xi)$ for $\xi\geq0$ is more complicated than that of $g(\xi)$. Because of the $-\ln\xi$ contribution, $h(\xi)$ becomes more and more positive when $\xi$ approaches zero. In fact, $h(\xi)$ strictly decreases from $+\infty$ to $-\infty$ for $\xi$ ranging from 0 to $+\infty$.

It is interesting to consider first the limiting case $b \stackrel{>}{\to} \frac{1}{2}$, yielding $\xi\approx 1.0612$.   So, there is a discontinuous jump in $\xi$ (i.e.~the Gribov parameter for fixed $v$) when the parameter $b$ crosses the boundary value $\frac{1}{2}$.

We were able to separate the $b>\frac{1}{2}$ region as follows:
\begin{itemize}
\item[(a)] For $\frac{1}{2}<b<\frac{1}{\sqrt{\sqrt{2}e}}\approx 0.51$, we have a unique solution $\xi>1$, i.e.~we are in the confining region again, with all gauge bosons displaying a Gribov type of propagator with complex conjugate poles.
\item[(b)] For $\frac{1}{\sqrt{\sqrt{2}e}}< b<\infty$, we have a unique solution $\xi<1$, indicating again a combination of two Yukawa modes for the off-diagonal gauge bosons. The ``photon'' is still of the Gribov type, thus confined.
\end{itemize}
Completely analogous as in the fundamental case, it can be checked by addressing the averages of the expressions \eqref{sigma} that for $b>\frac{1}{2}$ and $\omega$ obeying the gap equation with $\beta=0$, we are already within the Gribov horizon, making the introduction of the second Gribov parameter $\beta$ obsolete.

It is obvious that the transitions in the adjoint case are far more intricate than in the earlier studied fundamental case. First of all, we notice that the ``photon'' (diagonal gauge boson) is confined according to its Gribov propagator. There is never a Coulomb phase for $b<\infty$. The latter finding can be understood again from the viewpoint of the ghost self-energy. If the diagonal gluon would remain Coulomb (massless), the off-diagonal ghost self-energy, cfr.~eq.\eqref{sigma}, will contain an untamed infrared contribution from this massless photon\footnote{The ``photon'' indeed keeps it coupling to the charged (= off-diagonal) ghosts, as can be read off directly from the Faddeev-Popov term $c^a \p_\mu D_\mu^{ab}c^b$.}, leading to an off-diagonal ghost self-energy that will cross the value 1 at a momentum $k^2>0$, indicative of trespassing the first Gribov horizon. This crossing will not be prevented at any finite value of the Higgs condensate $\nu$, thus we are forced to impose at any time a nonvanishing Gribov parameter $\omega$. Treating the gauge copy problem for the adjoint Higgs sector will screen (rather confine) the a priori massless ``photon''.

An interesting limiting case is that of infinite Higgs condensate, also considered in the lattice study of \cite{Brower:1982yn}. Assuming $\nu\to\infty$, we have $b\to\infty$ according to its definition \eqref{vb}. Expanding the gap equation \eqref{gapddn} around $\xi=0^+$, we find the limiting equation $b^4=\frac{1}{\xi}$, or equivalently $\omega^*\propto \lms^8/g^4\nu^4$. Said otherwise, we find that also the second Gribov parameter vanishes in the limit of infinite Higgs condensate. As a consequence, the photon becomes truly massless in this limit. This result provides ---in our opinion--- a kind of continuum version of the existence of the Coulomb phase in the same limit as in the lattice version of the model probed in \cite{Brower:1982yn}. It is instructive to link this back to the off-diagonal no pole function, see Eq.~\eqref{sigma}, as we have argued in the proceeding paragraph that the massless photon leads to $\sigma_{off}(0)>1$ upon taking averages. However, there is an intricate combination of the limits $\nu\to\infty$, $\omega^*\to 0$ preventing such a problem here. Indeed, we find in these limits, again using dimensional regularization in the $\MSbar$ scheme, that
\begin{align}\label{dv}
  \sigma_{off}(0) &= \frac{3g^2}4 \left(\int \frac{d^4q}{(2\pi)^4} \frac1{q^4+\frac{\omega^\ast g^2}2} + \int \frac{d^4q}{(2\pi)^4} \frac1{q^2(q^2+g^2\nu^2)+\frac{\omega^\ast g^2}4}\right) \nonumber\\
	&= -\frac{3g^2}{128\pi^2} \left(\tfrac12 \ln\frac{\omega^\ast g^2}{2\overline\mu^4} + \ln\frac{g^2\nu^2}{\overline\mu^2}-2\right) \nonumber \\
	&= -\frac{3g^2}{128\pi^2} \left(\tfrac12 \ln\frac{\omega^\ast g^6\nu^4}{2\overline\mu^8}-2\right) \stackrel{b^4=\xi^{-1}}{\longrightarrow} -\frac{3g^2}{64\pi^2} \ln 8g^2 + \frac12.
\end{align}
The latter quantity is always smaller than $1$ for $g^2$ positive, meaning that we did not cross the Gribov horizon. This observation confirm in an explicit way the intuitive reasoning also found in section 3.4 of \cite{Lenz:2000zt}, at least in the limit $\nu\to\infty$. The subtle point in the above analysis is that it is not allowed to naively throw away the 2nd integral in the first line of \eqref{dv} for $\nu\to\infty$. There is a logarithmic $\ln\nu$ ($\nu\to\infty$) divergence that conspires with the $\ln \omega^*$ ($\omega^*\to0$) divergence of the 1st integral to yield the final reported result. This displays that, as usual, certain care is needed when taking infinite mass limits in Feynman integrals.

\subsubsection{The vacuum energy in the adjoint case}
As done in the case of the fundamental representation, let us work out the expression of the vacuum energy ${\cal E}_v$, for which we have the one loop  integral representation given by eq.\eqref{Zq4} multiplied by $-1$.
%\begin{eqnarray}
%{\cal E}_v= -\beta^*-\omega^*+\frac{3}{2}\int \frac{d^4q}{(2\pi)^4}\ln(q^4+2\tau')+3\int \frac{d^4q}{(2\pi)^4} \ln(q^4+g^2\nu^2q^2+\tau)
%\end{eqnarray}
%where
%\begin{equation}\label{ddvac1}
%    \tau'=\frac{g^2\omega^*}{4}\,,\qquad \tau=g^2\left(\frac{\beta^*}{2}+\frac{\omega^*}{4}\right).
%\end{equation}
%Based on \eqref{intdl} and on
%\begin{equation}\label{ddvac2}
%    \int\frac{d^4q}{(2\pi)^4}\ln(q^4+m^4)=-\frac{m^4}{32\pi^2}\left(\ln\frac{m^4}{\omu^4}-3\right),
%\end{equation}
%we find
Making use of the $\MSbar$ renormalization scheme in $d = 4 − \varepsilon$ the vacuum energy becomes
%\begin{eqnarray}\label{ddvac3}
%{\cal E}_v= -\frac{2}{g^2}(\tau+\tau')-\frac{3\tau'}{32\pi^2}\left(\ln\frac{2\tau'}{\omu^4}-3\right)+\frac{3}{32\pi^2}\left(m_-^4\left(\ln\frac{m_-^2}{\omu^2}-\frac{3}{2}\right)+q_+^4\left(\ln\frac{m_+^2}{\omu^2}-\frac{3}{2}\right)\right),
%\end{eqnarray}

%We can introduce $b$ via its definition \eqref{vb} to write after simplification
\begin{eqnarray}\label{ddvac4}
\frac{{\cal E}_v}{g^4\nu^4}&=& -\frac{1}{g^2}-\frac{3\xi'}{128\pi^2}\left(\ln(2b^2\xi')-1\right)+\frac{3(4-2\xi)}{128\pi^2}\left(\ln b-\frac{1}{2}\right)\nonumber\\
&&+\frac{3}{128\pi^2}\left(\left(1-\sqrt{1-\xi}\right)^2\ln\left(1-\sqrt{1-\xi}\right)+\left(1+\sqrt{1-\xi}\right)^2\ln\left(1+\sqrt{1-\xi}\right)\right)\;,
\end{eqnarray}
where $b$ was introduced via its definition \eqref{vb}, while
\begin{equation}\label{ddvac5}
    \xi'=\frac{4\tau'}{g^4\nu^4}\,,\qquad \xi=\frac{4\tau}{g^4\nu^4} 
\qquad \text{and} \qquad 
\tau'=\frac{g^2\omega^*}{4}\,,\qquad \tau=g^2\left(\frac{\beta^*}{2}+\frac{\omega^*}{4}\right)    \;.
\end{equation}

\begin{figure}[h!]
\center
%%\framebox[79mm]{\rule[-26mm]{0mm}{52mm}}
\includegraphics[width=8cm]{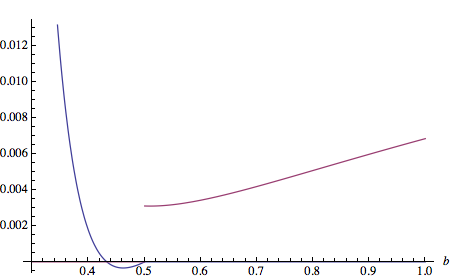}
\caption{Plot of the vacuum energy in the adjoint representation as a function of the parameter $b$. The discontinuity at $b=\frac{1}{2}$ is evident. }
\label{BS-d4}
\end{figure}

Since we found scenarios completely different for $b<1/2$ and $b>1/2$ with the scalar Higgs field in its adjoint representation, it becomes of great importance analysing the plot of the vacuum energy as a function of $b$. From Fig.\ref{BS-d4} one can easily find out a clear jump for $b=1/2$, which can be seen as a reflection of the discontinuity of the parameter  $\xi$.
%First, for $b<1/2$, we can use the  $\omega$-gap equation \eqref{ggapp1} to establish $\xi'=\frac{1}{2b^2}$, and thus
%\begin{eqnarray}\label{ddvac4bis}
%\frac{{\cal E}_v}{g^4\nu^4}&=& -\frac{1}{g^2}+\frac{3}{256b^2\pi^2}+\frac{3(4-2\xi)}{128\pi^2}\left(\ln b-\frac{1}{2}\right)\nonumber\\
%&&+\frac{3}{128\pi^2}\left(\left(1-\sqrt{1-\xi}\right)^2\ln\left(1-\sqrt{1-\xi}\right)+\left(1+\sqrt{1-\xi}\right)^2\ln\left(1+\sqrt{1-\xi}\right)\right)  \;,
%\end{eqnarray}
%where $\xi$ is determined by the equations  \eqref{gapdd},\eqref{gapddbis}.
%For $b>1/2$, we can remember that $\beta=0$ and thus $\xi=\xi'$, in which case we can easily obtain
%\begin{eqnarray}
%\frac{{\cal E}_v}{g^4 \nu^4} & = & - \left( \frac{1}{g^2} + \frac{3}{64\pi^2}  \right)   + \frac{3}{64\pi^2} \xi+ \frac{3}{32\pi^2} \frac{1}{4} (4-2\xi) \ln(b) \nonumber \\
%&+& \frac{3}{32\pi^2} \frac{1}{4} \left( \left(1+\sqrt{1-\xi}\right)^2\;\ln\left(1+\sqrt{1-\xi}\right) + \left(1-\sqrt{1-\xi}\right)^2\;\ln\left(1-\sqrt{1-\xi}\right) \right) \nonumber \\
%&-& \frac{3}{32\pi^2} \frac{1}{4} \left(2\xi \ln(\sqrt{\xi}) +2\xi \ln\sqrt{2} + 2\xi \ln(b)   \right)   \;, \label{Evv2}
%\end{eqnarray}
%with $\xi$ now given by eq.\eqref{gapddn}.
%It is worth noticing that the discontinuity in the parameter $\xi$ directly reflects itself in a discontinuity in the vacuum energy, as is clear from the plot of Fig.\ref{BS-d4}.

Investigating the functional \eqref{ddvac4} in terms of $\xi$ and $\xi'$, it is numerically (graphically) rapidly established there is always a solution to the gap equations $\frac{\p \mathcal{E}_v }{\p \xi}=\frac{\p \mathcal{E}_v }{\p \xi'}=0$ for $b<\frac{1}{2}$, but the solution $\xi^*$ is pushed towards the boundary $\xi=0$ if $b$ approaches $\frac{1}{2}$, to subsequently disappear for $b>\frac{1}{2}$ \footnote{The gap solutions correspond to a local maximum, as identified by analysing the Hessian matrix of 2nd derivatives.}. In that case, we are forced to return on our steps as in the fundamental case and conclude that $\beta=0$, leaving us with a single variable $\xi=\xi'$ and a new vacuum functional to extremize. There is, a priori, no reason for these 2 intrinsically different vacuum functionals be smoothly joined at $b=\frac{1}{2}$. This situation is clearly different from what happens when a potential has e.g.~2 different local minima with different energy, where at a first order transition the two minima both become global minima, thereafter changing their role of local vs.~global. Evidently, the vacuum energy does not jump since it is by definition equal at the transition.

Nevertheless, a completely analogous analysis as for the fundamental case will learn that $b=\frac{1}{2}$ is beyond the range of validity of our approximation\footnote{A little more care is needed as the appearance of two Gribov scales complicate the log structure. However, for small $b$ the Gribov masses will dominate over the Higgs condensate and we can take  $\omu$ of the order of the Gribov masses to control the logs and get a small coupling. For large $b$, we have $\beta^*=0$ and a small $\omega^*$: the first log will be kept small by its pre-factor and the other logs can be managed by taking $\omu$ of the order of the Higgs condensate.}. The small and large $b$ results can again be shown to be valid, so at large $b$ ($\sim$ large Higgs condensate) we have a mixture of off-diagonal Yukawa and confined diagonal modes and at small $b$ ($\sim$ small Higgs condensate) we are in a confined phase. In any case we have that the diagonal gauge boson is \emph{not} Coulomb-like, its infrared behaviour is suppressed as it feels the presence of the Gribov horizon.

%%%%%%%%%%%%%%%%%%%%%%%%%%%%%%%%%%%%%%%%%%%%%%%%%%%%%%%%%%%%%%%%
\section{The Electroweak theory: $SU(2)\times U(1)+$Higgs field}
\label{The Electroweak theory}
%%%%%%%%%%%%%%%%%%%%%%%%%%%%%%%%%%%%%%%%%%%%%%%%%%%%%%%%%%%%%%%%

From now on in this work only the fundamental case of the Higgs field will be treated, for reasons relying on the physical relevance of the fundamental representation of this field. As a first step, we are going to present, as in the previous sections, general results for $d$-dimension. Afterwards, the $3$ and $4$-dimensional cases will be considered in the subsections \ref{d=3} and \ref{d=4}. The starting action of the $SU(2) \times U(1)+$Higgs field reads
\begin{eqnarray}
S ~=~ \int \d^{d}x  \;  \bigg(\frac{1}{4}  F_{\mu \nu }^{a} F_{\mu \nu }^{a}  +  \frac{1}{4} B_{\mu\nu} B_{\mu\nu} +{\bar c}^a \partial_\mu D^{ab}_\mu c^b - \frac{(\partial_\mu A^a_\mu)^{2}}{2\xi} 
 + {\bar c}\partial^2 c  - \frac{(\partial_\mu B_\mu)^{2}}{2\xi}  +
\nonumber \\
+
(D_{\mu }^{ij}\Phi^{j})^{\dagger}( D_{\mu }^{ik}\Phi^{k})+\frac{\lambda }{2}\left(\Phi^{\dagger}\Phi - \nu^{2}\right)^{2}   \bigg)  \;,
\label{Sf}
\end{eqnarray}
where the covariant derivative is defined by
\begin{equation}
D_{\mu }^{ij}\Phi^{j} =\partial _{\mu }\Phi^{i} - \frac{ig'}{2}B_{\mu}\Phi^{i} -   ig \frac{(\tau^a)^{ij}}{2}A_{\mu }^{a}\Phi^{j}  \;.
\end{equation}
and the vacuum expectation value (\textit{vev}) of the Higgs field is $\langle \Phi^{i} \rangle ~=~ \nu\delta^{2i}$.
%%\begin{equation}
%\langle \Phi \rangle  = \left( \begin{array}{ccc}
%                                          0  \\
%                                          \nu
%                                          \end{array} \right)  \;.
%\label{vevf}
%%\end{equation}
The indices $i,j=1,2$ refer to the fundamental representation of $SU(2)$ and $\tau^a, a=1,2,3$, are the Pauli matrices. The coupling constants $g$ and $g'$ refer to the groups $SU(2)$ and $U(1)$, respectively. The field strengths $F^a_{\mu\nu}$ and $B_{\mu\nu}$ are given by
\begin{equation}
F^a_{\mu\nu} = \partial_\mu A^a_\nu -\partial_\nu A^a_\mu + g \varepsilon^{abc} A^b_\mu A^c_\nu \;, \qquad B_{\mu \nu} = \partial_\mu B_\nu -\partial_\nu B_\mu  \;.
\label{fs}
\end{equation}

In order to obtain the boson propagators only quadratic terms of the starting action are required and, due to the new covariant derivative, this quadratic action is not diagonal any more. To diagonalize this action one could introduce a set of new fields, related to the standard ones by
%To obtain the  gauge boson propagators, we consider the quadratic part of the action (\ref{Sf}), given by
%\begin{multline}
%S_{quad} = \int\!\! \d^{d}x \; \frac{1}{2}A_{\mu}^{\alpha}\left[ \left(-\partial_{\mu}\partial_{\mu} + \frac{\nu^{2}}{2}g^{2}\right)\delta_{\mu\nu} + \partial_{\mu}\partial_{\nu} \right]A_{\nu}^{\alpha} + 
%\int\!\! \d^{d}x\, \frac{1}{2}B_{\mu}\left[ \left(-\partial_{\mu}\partial_{\mu} + \frac{\nu^{2}}{2}g'^{2}\right)\delta_{\mu\nu} + \partial_{\mu}\partial_{\nu}\right]B_{\nu} \\
%+ \int\!\! \d^{d}x\, \frac{1}{2}A_{\mu}^{3}\left[ \left(-\partial_{\mu}\partial_{\mu} + \frac{\nu^{2}}{2}g^{2}\right)\delta_{\mu\nu} + \partial_{\mu}\partial_{\nu}\right]A_{\nu}^{3} - \frac{1}{4}
%\int\!\! \d^{d}x\, \nu^{2}g\,g'\,A^{3}_{\mu}B_{\mu} - \frac{1}{4} 
%\int\!\! \d^{d}x\, \nu^{2}g\,g'\,B_{\mu}A^{3}_{\mu} \;.
%\label{Squad}
%\end{multline}
%In order to diagonalize expression \eqref{Squad} we introduce the following fields
\begin{subequations} \begin{gather}
W^+_\mu = \frac{1}{\sqrt{2}} \left( A^1_\mu + iA^2_\mu \right) \;, \qquad W^-_\mu = \frac{1}{\sqrt{2}} \left( A^1_\mu - iA^2_\mu \right)  \;,
\label{ws} \\
Z_\mu =\frac{1}{\sqrt{g^2+g'^2} } \left(  -g A^3_\mu + g' B_\mu \right) \qquad \text{and}\qquad A_\mu =\frac{1}{\sqrt{g^2+g'^2} } \left(  g' A^3_\mu + gB_\mu \right) \;.
\label{za}
\end{gather} 
\end{subequations}
The inverse relation can be easily obtained.
%Let us also give, for further use, the inverse combinations:
%\begin{subequations} \begin{gather}
%A^1_\mu = \frac{1}{\sqrt{2}} \left( W^+_\mu + W^-_\mu \right) \;, \qquad A^2_\mu = \frac{1}{i\sqrt{2}} \left( W^+_\mu - W^-_\mu \right) \;,
%\label{iw} \\
%B_\mu =\frac{1}{\sqrt{g^2+g'^2} } \left(  g A_\mu + g' Z_\mu \right) \qquad \text{and}\qquad A^3_\mu =\frac{1}{\sqrt{g^2+g'^2} } \left(  g' A_\mu - gZ_\mu \right) \;.
%\label{iza}
%\end{gather} \end{subequations}
With this new set of fields the quadratic part of the action reads,
\begin{eqnarray}
S_{quad} &=&  \int d^3 x   \left( \frac{1}{2} (\partial_\mu W^+_\nu - \partial_\nu W^+_\mu)(\partial_\mu W^-_\nu - \partial_\nu W^-_\mu)  + \frac{g^2\nu^2}{2}W^+_\mu W^-_\mu   \right) 
\nonumber \\
&+& \int d^3x  \left(  \frac{1}{4} (\partial_\mu Z_\nu - \partial_\nu Z_\mu)^2  + \frac{(g^2+g'^2)\nu^2}{4}Z_\mu Z _\mu  +    \frac{1}{4} (\partial_\mu A_\nu - \partial_\nu A_\mu)^2  \right)  \;,
\label{qd}
\end{eqnarray}
from which we can read off the masses of the fields $W^+$, $W^-$, and $Z$:
\begin{equation}
m^2_W = \frac{g^2\nu^2}{2} \;, \qquad m^2_Z =  \frac{(g^2+g'^2)\nu^2}{2}  \;. \label{ms}
\end{equation}

The restriction to the Gribov region $\Omega$ still is needed and the procedure here becomes quite similar to what was carried out in the subsection \ref{Adjrep}. Due to the breaking of the global gauge invariance, caused by the Higgs field (through the covariant derivatives), the ghost sector can be split up in two different sectors, diagonal and off-diagonal. Namely, the ghost propagator reads,

\begin{equation}
\mathcal{G}^{ab}(k;A) ~=~ \left(
  \begin{array}{ll}
   \delta^{\alpha \beta} \mathcal{G}_{off}(k;A) & \,\,\,\,\,\,\,\,0 \\
   \,\,\,\;\;\;\;\;\;0 & \mathcal{G}_{diag}(k;A)
  \end{array}
\right).
\label{gh prop offdiag}
\end{equation}
By expliciting the ghost form factor we have
\begin{eqnarray}
\mathcal{G}_{off}(k;A) 
~\simeq~  \frac{1}{k^{2}} \left( \frac{1}{1 - \sigma_{off}(k;A)} \right) \;,
\label{gh off}
\end{eqnarray}
and
\begin{eqnarray}
\mathcal{G}_{diag}(k;A) 
~\simeq ~ \frac{1}{k^{2}} \left( \frac{1}{1 - \sigma_{diag}(k;A)} \right)\;,
\label{gh diag}
\end{eqnarray}
where
\begin{subequations} \begin{equation}
\sigma_{off}(0;A) ~=~ \frac{g^{2}}{dV} \int\!\! \frac{\d^{d}p}{(2\pi)^{d}} \;  \frac{1}{p^{2}} \left( \frac{1}{2} A^{\alpha}_{\mu}(p)A^{\alpha}_{\mu}(-p) + A^{3}_{\mu}(p)A^{3}_{\mu}(-p)\right)  \;,
\label{sigma off}
\end{equation}
and
\begin{equation}
\sigma_{diag}(0;A) ~=~ \frac{g^{2}}{dV} \int\!\! \frac{\d^{d}p}{(2\pi)^{d}} \; \frac{1}{p^{2}} A^{\alpha}_{\mu}(p)A^{\alpha}_{\mu}(-p)\;.
\label{sigma diag}
\end{equation} \end{subequations}
In order to obtain expressions  \eqref{sigma off} and \eqref{sigma diag}, where $V$ denotes the (infinte) space-time volume, the transversality of the gluon field and the property that $\sigma(k;A)_{off}$ and $\sigma(k;A)_{diga}$ are decreasing functions of $k$ were used\footnote{For more details concerning the ghost computation see \cite{Capri:2013oja,Capri:2013gha,Capri:2012ah,Vandersickel:2012tz}}. From equations \eqref{gh off} and \eqref{sigma off} one can easily read off the two no-pole conditions. Namely,
\begin{subequations} \begin{equation}
\sigma_{off}(0;A) < 1  \;,
\label{sigmaoffnopole}
\end{equation}
and
\begin{equation}
\sigma_{diag}(0;A) < 1\;.
\label{sigmadiagnopole}
\end{equation} \end{subequations}

%Expressions  \eqref{sigma off} and \eqref{sigma diag} are obtained by taking the limit $k \to 0$ of equations \eqref{gh off},  \eqref{gh diag}, and by making use of the property
%\begin{eqnarray}
%A_{\mu }^{a}(q)A_{\nu }^{a}(-q) &=&\left( \delta _{\mu \nu }-\frac{q_{\mu }q_{\nu }}{q^{2}}\right) \omega (A)(q)  
%\nonumber \\
%&\Rightarrow &\omega (A)(q) ~=~ \frac{1}{(d-1)}A_{\lambda }^{a}(q)A_{\lambda }^{a}(-q)    \;,
%\label{p1}
%\end{eqnarray}
%which follows from the transversality of the gauge field, $q_\mu A^a_\mu(q)=0$. Also, it is useful to remind that, for an arbitrary function $\mathcal{F}(p^2)$, we have
%\begin{equation}
%\int \frac{\d^{d}p}{(2\pi )^{d}}\left( \delta _{\mu \nu }-\frac{p_{\mu }p_{\nu }}{p^{2}}\right) \mathcal{F}(p^2) ~=~ \mathcal{A}\;\delta _{\mu \nu }   \;,
%\label{p2}
%\end{equation}
%where, upon contracting both sides of equation \eqref{p2} with $\delta_{\mu\nu}$,
%\begin{equation}
%\mathcal{A} ~=~ \frac{d-1}{d}\int \frac{\d^{d}p}{(2\pi )^{d}}\mathcal{F}(p^2)\;.   
%\label{p3}
%\end{equation}

At the end, the partition function restricted to the first Gribov region $\Omega$ reads,
\begin{eqnarray}
Z &=&  \int \frac{\d \omega}{2\pi i \omega}\frac{\d \beta}{2\pi i \beta} [\d A] [\d B]  \; \; e^{\omega (1-\sigma_{off})} \, e^{\beta (1-\sigma_{diag})} e^{-S}\;.
\label{ptionfucnt22}
\end{eqnarray}

\subsubsection{The $d$-dimensional gluon propagator and the gap equation}

The perturbative computation at the semi-classical level requires only quadratic terms of the full action, defined in eq.\eqref{ptionfucnt22} (with $S$ given by eq.\eqref{Sf}), yielding a Gaussian integral over the fields. Inserting external fields to obtain the boson propagators, one gets, after taking the limit $\xi \to 0$, the following propagators,

\begin{subequations} \label{propsaandb} \begin{gather}
\langle  A^{\alpha}_\mu(p) A^{\beta}_\nu(-p) \rangle ~=~ \frac{p^2}{p^4 + \frac{\nu^2g^2}{2} p^2 + \frac{2g^2\beta}{dV}} \; \delta^{\alpha \beta} \left( \delta_{\mu\nu} - \frac{p_\mu p_\nu}{p^2} \right)  \;,   
\label{aalpha} \\
\langle  A^{3}_\mu(p) A^{3}_\nu(-p) \rangle ~=~ \frac{p^2 \left(p^2 +\frac{\nu^2}{2} g'^{2}\right)}{p^6 + \frac{\nu^2}{2} p^4 \left(g^2 +g'^2 \right)  + \frac{2\omega g^2}{dV} \left( p^2 + \frac{\nu^2 g'^2}{2} \right)} \;  \left( \delta_{\mu\nu} - \frac{p_\mu p_\nu}{p^2} \right)  \;,   \label{a3a3}
\\ %
\langle  B_\mu(p) B_\nu(-p) \rangle ~=~ \frac{ \left(p^4 +\frac{\nu^2}{2} g^{2} p^2+\frac{2\omega g^2}{dV}  \right)}{p^6 + \frac{\nu^2}{2} p^4 \left(g^2 +g'^2 \right)  + \frac{2\omega g^2}{dV} \left( p^2 +  \frac{\nu^2 g'^2}{2} \right)} \;  \left( \delta_{\mu\nu} - \frac{p_\mu p_\nu}{p^2} \right)  \;,   \label{bb}
\\
\langle  A^3_\mu(p) B_\nu(-p) \rangle ~=~  \frac{ \frac{\nu^2}{2} g g'  p^2}{p^6 + \frac{\nu^2}{2} p^4 \left(g^2 +g'^2 \right)  + \frac{2\omega g^2}{dV} \left( p^2 + \frac{\nu^2 g'^2}{2} \right)} \;  \left( \delta_{\mu\nu} - \frac{p_\mu p_\nu}{p^2} \right)  \;.   \label{ba3}
\end{gather} \end{subequations}
Moving to the fields $W^{+}_\mu, W^{-}_\mu, Z_\mu, A_\mu$, one obtains 
\begin{subequations} \label{propszandgamma} \begin{gather}
\langle  W^{+}_\mu(p) W^{-}_\nu(-p) \rangle ~=~ \frac{p^2}{p^4 + \frac{\nu^2g^2}{2} p^2 + \frac{2g^2\beta}{dV} } \;  \left( \delta_{\mu\nu} - \frac{p_\mu p_\nu}{p^2} \right)  \;,   \label{ww}
\\
\langle  Z_\mu(p) Z_\nu(-p) \rangle ~=~ \frac{\left( p^4 +\frac{2\omega}{dV} \frac{g^2 g'^2}{g^2+g'^2}  \right)}{p^6 + \frac{\nu^2}{2} p^4 \left(g^2 +g'^2 \right)  + \frac{2\omega g^2}{dV} \left( p^2 + \frac{\nu^2 g'^2}{2} \right) } \;  \left( \delta_{\mu\nu} - \frac{p_\mu p_\nu}{p^2} \right)  \;,   \label{zz}
\\
\langle  A_\mu(p) A_\nu(-p) \rangle ~=~ \frac{\left( p^4 +\frac{\nu^2}{2} p^2 (g^2+g'^2) +\frac{2\omega}{dV} \frac{g^4}{g^2+g'^2}\right)}{p^6 + \frac{\nu^2}{2} p^4 \left(g^2 +g'^2 \right)  + \frac{2\omega g^2}{dV} \left( p^2 + \frac{\nu^2 g'^2}{2} \right) } \;  \left( \delta_{\mu\nu} - \frac{p_\mu p_\nu}{p^2} \right)  \;,   \label{aa}
\\
\langle  A_\mu(p) Z_\nu(-p) \rangle ~=~ \frac{\frac{2\omega}{dV} \frac{g^3 g'}{g^2+g'^2} }{p^6 + \frac{\nu^2}{2} p^4 \left(g^2 +g'^2 \right)  + \frac{2\omega g^2}{dV} \left( p^2 + \frac{\nu^2 g'^2}{2} \right)} \;  \left( \delta_{\mu\nu} - \frac{p_\mu p_\nu}{p^2} \right)  \;.   \label{az}
\end{gather} \end{subequations}
As expected, all propagators get deeply modified in the IR by the presence of the Gribov parameters $\beta$ and $\omega$. Notice, in particular, that due to the parameter $\omega$ a mixing between the fields $A_\mu$ and $Z_\mu$ arises, eq.\eqref{az}. As such, the original photon and the boson $Z$ loose their distinct particle interpretation.  Moreover, it is straightforward to check that in the limit $\beta \rightarrow 0$ and $\omega \rightarrow 0$, the standards propagators are recovered.
%, {\it i.e.}
%\begin{subequations} \begin{gather}
%\langle  W^{+}_\mu(p) W^{-}_\nu(-p) \rangle {\big |}_{\beta=0} ~=~ \frac{1}{p^2 + \frac{\nu^2g^2}{2} } \;  \left( \delta_{\mu\nu} - \frac{p_\mu p_\nu}{p^2} \right)  \;,   \label{ww0}
%\\
%\langle  Z_\mu(p) Z_\nu(-p) \rangle  {\big |}_{\omega=0} ~=~ \frac{1}{p^2 + \frac{\nu^2}{2} \left(g^2 +g'^2 \right)} \;  \left( \delta_{\mu\nu} - \frac{p_\mu p_\nu}{p^2} \right)  \;,   \label{zz0}
%\\
%\langle  A_\mu(p) A_\nu(-p) \rangle{\big |}_{\omega=0}  ~=~ \frac{1}{p^2}   \;  \left( \delta_{\mu\nu} - \frac{p_\mu p_\nu}{p^2} \right)  \;,   \label{aa0}
%\\
%\langle  A_\mu(p) Z_\nu(-p) \rangle {\big |}_{\omega=0} = 0    \;.   \label{az0}
%\end{gather} 
%\end{subequations}

Let us now proceed by deriving the gap equations which will enable us to (dynamically) fix the Gribov parameters, $\beta$ and $\omega$, as function of $g$, $g'$ and $\nu^2$. Thus, performing the path integral of eq.\eqref{ptionfucnt22}, in the semi-classical level, we get
%\begin{equation}
%Z_{quad} ~=~ \mathcal{N'} \int \frac{\d \omega}{2\pi i}\frac{\d \beta}{2\pi i}\,e^{-\ln\left(\beta\omega -\frac{\omega^{2}}{2}\right)} e^{\frac{\omega}{2}}\,e^{\beta} \left[\det Q_{\mu \nu}^{\alpha \beta} \right]^{-1/2}\, \left[ \det \mathcal{P}_{\mu \nu} \right]^{-1/2},  \label{Zquad3}
%\end{equation}
%with
%\begin{subequations} \begin{equation}
%\left[\det Q_{\mu \nu}^{\alpha \beta} \right]^{-1/2} ~=~ \exp \left[ -\frac{2(d-1)}{2} \int \frac{\d^{d}p}{(2\pi)^{d}}  \;  \log \left(p^{2} + \frac{g^{2}\nu^{2}}{2} + \frac{2g^{2}\beta}{dV} \frac{1}{p^{2}}\right) \right] \label{calc_det1}
%\end{equation}
%and
%\begin{equation}
%\left[ \det \mathcal{P}_{\mu \nu} \right]^{-1/2} ~=~ \exp \left[ - \frac{(d-1)}{2} \int \frac{\d^{d}p}{(2\pi)^{d}} \log \lambda_{+}(p, \omega)\, \lambda_{-}(p, \omega) \right].   \label{calc_det2}
%\end{equation} \end{subequations}
%where $\lambda_{\pm}$ are the eigenvalues of the $2\times 2$ matrix of eq.\eqref{P mu nu}, {\it i.e.}
%\begin{equation}
%\lambda_{\pm} ~=~ \frac{\left( p^{4} + \frac{\nu^{2}}{4} p^{2}(g^{2} + g'^{2}) + \frac{g^{2}\omega}{dV} \right) \pm \sqrt{\left[ \frac{\nu^{2}}{4}(g^{2} + g'^{2})p^{2} + \frac{g^{2}\omega}{dV}\right]^{2} - \frac{\omega}{3}\nu^{2}g^{2}\,g'^{2}p^{2}}}{p^{2}}  \;. \label{ev}
%\end{equation}
%Thus,
%\begin{equation}
%Z_{quad} = \mathcal{N} \int \frac{\d \omega}{2\pi i}\frac{\d \beta}{2\pi i} e^{f(\omega, \beta)} \;, \label{Zf eq}
%\end{equation}
%where
\begin{eqnarray}
f(\omega, \beta) ~=~ \frac{\omega}{2} + \beta - \frac{2(d-1)}{2} \int \frac{\d^{d}p}{(2\pi)^{d}} \; \log \left(p^{2} + \frac{\nu^{2}}{2}g^{2} + \frac{2g^{2}\beta}{dV} \frac{1}{p^{2}}\right) -
\nonumber \\
- \frac{(d-1)}{2} \int \frac{\d^{d}p}{(2\pi)^{d}} \; \log \lambda_{+}(p, \omega)\, \lambda_{-}(p, \omega)\;. 
\label{f eq}
\end{eqnarray}
In eq.\eqref{f eq}, $f(\omega,\beta)$ is defined according to eq.\eqref{Zq3} and
\begin{equation}
\lambda_{\pm} ~=~ \frac{\left( p^{4} + \frac{\nu^{2}}{4} p^{2}(g^{2} + g'^{2}) + \frac{g^{2}\omega}{dV} \right) \pm \sqrt{\left[ \frac{\nu^{2}}{4}(g^{2} + g'^{2})p^{2} + \frac{g^{2}\omega}{dV}\right]^{2} - \frac{\omega}{3}\nu^{2}g^{2}\,g'^{2}p^{2}}}{p^{2}}  \;. 
\label{ev}
\end{equation}
Making use of the thermodynamic limit, where the saddle point approximation takes place, we have the two gap equations given by\footnote{For more details see \cite{Capri:2013oja,Capri:2013gha,Capri:2012ah}.}

\begin{equation}
\frac{4(d-1)}{2d}g^{2} \int \frac{\d^{d}p}{(2\pi)^{d}} \;  \frac{1}{p^{4}+\frac{g^{2}\nu^{2}}{2}p^{2} + \frac{2g^{2}\beta^{\ast}}{dV} } ~=~ 1   \;,
\label{beta gap eq}
\end{equation} 
and
%In particular, after a little algebra, eq.\eqref{omega gap eq1} can be considerably simplified, yielding
\begin{equation}
\frac{2(d-1)}{d}g^{2}\int\!\! \frac{\d^{d}p}{(2\pi)^{d}} \; \frac{p^{2} + \frac{\nu^{2}}{2}g'^{2}}{p^{6} + \frac{\nu^{2}}{2}(g^{2} + g'^{2})p^{4} + \frac{2\omega^{\ast}g^{2}}{dV}p^{2} + \frac{\nu^{2}g^{2}\,g'^{2}\omega^{\ast}}{dV} } ~=~ 1 \;.
\label{omega gap eq}
\end{equation}

Given the difficulties in solving the gap equations \eqref{beta gap eq} and \eqref{omega gap eq}, we propose an alternative approach to probe the gluon propagators in the parameter space $\nu$, $g$ and $g'$. Instead of explicitly solve the gap equations, let us search for the necessity to implement the Gribov restriction. For that we mean to compute $\langle \sigma_{off}(0) \rangle $ and $\langle \sigma_{diag}(0) \rangle$ with the gauge field propagators unchanged by the Gribov terms, {\it i.e.}, before applying the Gribov restriction. Therefore, if $\langle \sigma_{off}(0;A) \rangle  < 1$ and $\langle \sigma_{diag}(0;A) \rangle < 1$ already in this case (without Gribov restrictions), then we would say that there is no need to restrict the domain of integration to $\Omega$. In that case we have, immediately, $\beta^* = \omega^* =0$ and the standard Higgs procedure takes place. Namely, the expression of each ghost form factor is
\begin{eqnarray}
\langle \sigma_{off}(0) \rangle & = &  \frac{(d-1)g^{2}}{d}  \int\!\!  \frac{\d^d p}{(2\pi)^d} \frac{1}{p^{2}}\left(\frac{1}{p^{2} + \frac{\nu^{2}}{2}g^{2}} + \frac{1}{p^{2} + \frac{\nu^{2}}{2}(g^{2}+g'^{2})} \right)  \;.
\label{sgoff1}
\end{eqnarray}
and
\begin{equation}
\langle \sigma_{diag}(0) \rangle ~=~ \frac{2(d-1)g^{2}}{d} \int\!\!\frac{\d^{d}p}{(2\pi)^{d}}\frac{1}{p^{2}}\left(\frac{1}{p^{2} + \frac{\nu^{2}}{2}g^{2}}\right)  \;.
\label{sgdiag1}
\end{equation}

%\subsubsection{About $\sigma_\text{off}(0)$ and $\sigma_\text{diag}(0)$ without the Gribov parameters}
%\label{Evaluation of the ghost form factors}

%Specifically in the next two sections we propose a different approach to check if there exist values of the Higgs condensate $\nu$ and of the coupling $(g,\;g')$ for which both $\langle \sigma_{off}(0) \rangle $ and $\langle \sigma_{diag}(0) \rangle$  already satisfy the no-pole condition
%\begin{equation}
%\langle \sigma_{off}(0;A) \rangle  < 1  \;, \qquad   \langle \sigma_{diag}(0;A) \rangle < 1\;,
%\label{n-pole}
%\end{equation}
%in which case $\beta^\ast$ and/or $\omega^\ast$ could be immediately set equal to zero. Therefore, we present here the expression of both quantities for the $d$-dimensional case. Thus,

%Analogously, for $\langle \sigma_{diag}(0)\rangle$ one gets

%Now that we have in hands all expressions needed to analyze the gluon propagator, let us specialize them to the $d=3$ and $d=4$ cases. With that, we are able to provide a map of the parameter space displaying regions where we have sign of confinement, regions where the Higgs mechanism takes place unaltered and mixed regions where the propagators have ``confined'' and ``deconfined'' contributions in its expressions.

\subsection{The $d=3$ case} 
\label{d=3}

\begin{figure}\begin{center}
\includegraphics[width=.25\textwidth]{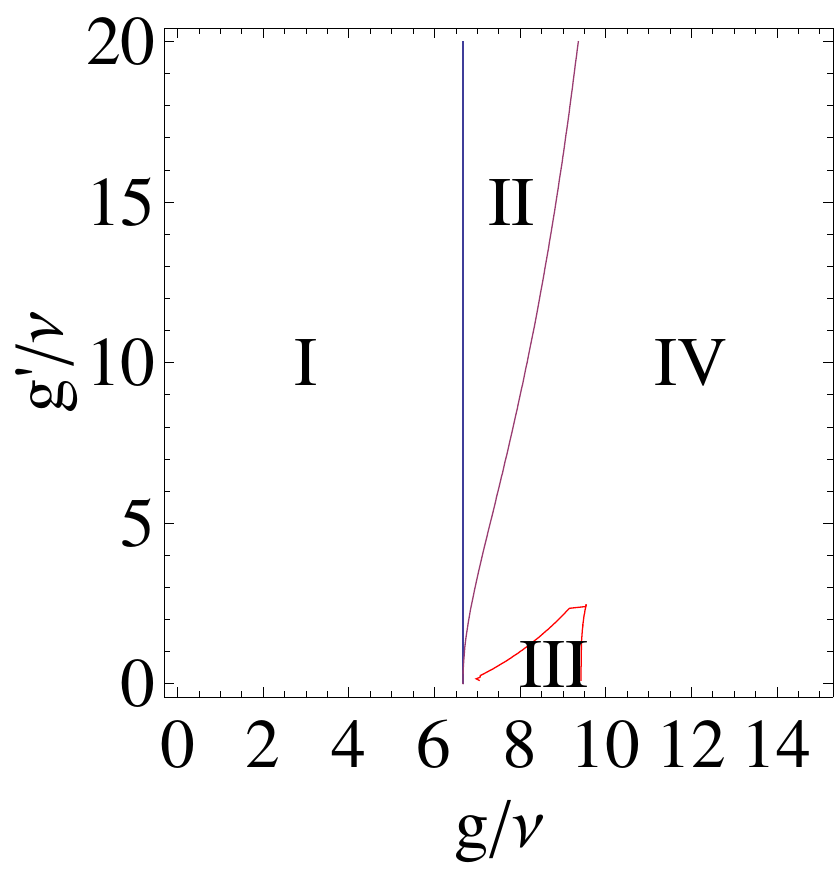}
\caption{There appear to be four regions in phase space. The region I is defined by condition \eqref{secondcond} and is characterized by ordinary Yang--Mills--Higgs behaviour (massive $W$ and $Z$ bosons, massless photon). The region II is defined by \eqref{firstcond} while excluding all points of region I --- this region only has electrically neutral excitations, as the $W$ bosons are confined (see Section \ref{sect6}); the massive $Z$ and the massless photon are unmodified from ordinary Yang--Mills--Higgs behaviour. Region III has confined $W$ bosons, while both photon and $Z$ particles are massive due to influence from the Gribov horizon; furthermore there is a negative-norm state. In region IV all $SU(2)$ bosons are confined and only a massive photon is left. Mark that the tip of region III is hard to deal with numerically --- the discontinuity shown in the diagram is probably an artefact due to this difficulty.  Details are collected in Section \ref{sect7}. \label{regionsdiag1}}
\end{center}\end{figure}

In the three-dimensional case things become easier since there is no divergences to treat. Therefore, computing the ghost form factors \eqref{sgoff1} and \eqref{sgdiag1} we led to the following conditions

\begin{subequations} 
\label{conds} 
\begin{eqnarray}
(1+\cos(\theta_{W}))\frac{g}{\nu} &<& 3\sqrt{2}\pi \label{firstcond} \\
2\frac{g}{\nu} &<& 3\sqrt{2}\pi \label{secondcond} \;,
\end{eqnarray} 
\end{subequations}
where $\theta(W)$ stands for the Weinberg angle,
\begin{equation}
\cos(\theta_{W}) = \frac{g}{\sqrt{g^{2}+g'^{2}}}\;.
\label{thW}
\end{equation}
These two conditions make phase space fall apart in three regions, as depicted in \figurename\ \ref{regionsdiag1}.
\begin{itemize}
	\item If $g/\nu<3\pi/\sqrt2$, neither Gribov parameter is necessary to make the integration cut off at the Gribov horizon. In this regime the theory is unmodified from the usual perturbative electroweak theory.
	\item In the intermediate case $3\pi/\sqrt2<g/\nu<3\sqrt2\pi/(1+\cos\theta_W)$ only one of the two Gribov parameters,  $\beta$, is necessary. The off-diagonal ($W$) gauge bosons will see their propagators modified due to the presence of a non-zero $\beta$, while the $Z$ boson and the photon $A$ remain untouched.
	\item In the third phase, when $g/\nu>3\sqrt2\pi/(1+\cos\theta_W)$, both Gribov parameters are needed, and all propagators are influenced by them. The off-diagonal gauge bosons are confined. The behaviour of the diagonal gauge bosons depends on the values of the couplings, and the third phase falls apart into two parts, as detailed in section \ref{zandgamma}.
\end{itemize}
Note that here in the $3$-dimensional $SU(2)\times U(1)+$Higgs case, as well as in the $3d$ $SU(2)+$Higgs treated in section \ref{3dsu2}, an effective coupling constant becomes of utmost importance when discussing the trustworthiness of the our semi-classical results.

\subsubsection{The off-diagonal ($W$) gauge bosons} 
\label{sect6}
Let us first look at the behaviour of the off-diagonal bosons under the influence of the Gribov horizon. The propagator \eqref{ww}  only contains the $\beta$ Gribov parameter, meaning that $\omega$ need not be considered here.

In the regime $g/\nu<3\pi/\sqrt2$ (region I in \figurename\ \ref{regionsdiag1}) the parameter $\beta$ is not necessary, due to the ghost form factor $\langle\sigma_{diag}(0)\rangle$ always being smaller than one. In this case, the off-diagonal boson propagator is simply of massive type, with mass parameter $\frac{\nu^{2}}{2}g^{2}$.
%\begin{equation}
%  \langle W^{+}_{\mu}(p)W^{-}_{\nu}(-p) \rangle = \frac{1}{p^{2} + \frac{\nu^{2}}{2}g^{2}}\left(\delta_{\mu\nu} - \frac{p_{\mu}p_{\nu}}{p^{2}}\right) \;.
%\end{equation}

In the case that $g/\nu>3\pi/\sqrt2$ (regions II, III, and IV in \figurename\ \ref{regionsdiag1}), the relevant ghost form factor is not automatically smaller than one any more, and the Gribov parameter $\beta$ becomes necessary. The value of $\beta^{\ast}$ is determined from the gap equations \eqref{beta gap eq}. After rewriting the integrand in partial fractions, the integral in the equation becomes of standard type, and we readily find the solution
\begin{equation}
  \beta^{\ast} = \frac{3g^2}{32} \left(\frac{g^2}{2\pi^2}-\nu^2\right)^2 \;.
\end{equation}
Mark that, in order to find this result, we had to take the square of both sides of the equation twice. One can easily verify that, in the region $g/\nu>3\pi/\sqrt2$ which concerns us, no spurious solutions were introduced when doing so.

Replacing this value of $\beta^{\ast}$ in the off-diagonal propagator \eqref{ww} one can immediately check that it
%\begin{multline}
%  \langle W_\mu^{+}(p)W_\nu^{-}(-p)\rangle = \frac{\pi/g^3}{\sqrt{\frac{g^2}{4\pi^2}-\nu^2}} \left(\frac{\frac{g^3}{2\pi}\sqrt{\frac{g^2}{4\pi^2}-\nu^2}-\frac i4\nu^2g^2}{p^2+\frac{\nu^2}4g^2+i\frac{g^3}{2\pi}\sqrt{\frac{g^2}{4\pi^2}-\nu^2}} + \frac{\frac{g^3}{2\pi}\sqrt{\frac{g^2}{4\pi^2}-\nu^2}+\frac i4\nu^2g^2}{p^2+\frac{\nu^2}4g^2-i\frac{g^3}{2\pi}\sqrt{\frac{g^2}{4\pi^2}-\nu^2}}\right) \\ \times \left(\delta_{\mu\nu}-\frac{p_\mu p_\nu}{p^2}\right) \;.
%\end{multline}
clearly displays two complex conjugate poles. As such, the off-diagonal propagator  cannot describe a physical excitation of the physical spectrum, being adequate for a confining phase. This means that the off-diagonal components of the gauge field are confined in the region $g/\nu>3\pi/\sqrt2$.

\subsubsection{The diagonal $SU(2)$ boson and the photon field} \label{zandgamma} \label{sect7}
The other two gauge bosons --- the $A^3_\mu$ and the $B_\mu$ --- have their propagators given by \eqref{a3a3}, \eqref{bb}, and \eqref{ba3} or equivalently --- the $Z_\mu$ and the $A_\mu$ --- by \eqref{zz}, \eqref{aa} and \eqref{az}. Here, $\omega$ is the only one of the two Gribov parameters present.

In the regime $g/\nu<3\sqrt2\pi/(1+\cos\theta_W)$ (regions I and II) this $\omega$ is not necessary to restrict the region of integration to within the first Gribov horizon. Due to this, the propagators are unmodified in comparison to the perturbative case.
%\begin{subequations} \label{propsrewrite} \begin{gather}
%\langle Z_{\mu}(p) Z_{\nu}(-p) \rangle = \frac{1}{p^{2} + \frac{\nu^{2}}{2}(g^{2} + g'^{2})} \left(\delta_{\mu\nu} - \frac{p_{\mu}p_{\nu}}{p^{2}}\right)\;, \\
%\langle A_{\mu}(p) A_{\nu}(-p) \rangle = \frac{1}{p^{2}} \left(\delta_{\mu\nu} - \frac{p_{\mu}p_{\nu}}{p^{2}}\right) \;.
%\end{gather} \end{subequations}

In the region $g/\nu>3\sqrt2\pi/(1+\cos\theta_W)$ (regions III and IV) the Gribov parameter $\omega$ does become necessary, and it has to be computed by solving its gap equation, eq. \eqref{omega gap eq}. Due to its complexity it seems impossible to do so analytically. Therefore we turn to numerical methods. Using Mathematica the gap equation can be straightforwardly solved for a list of values of the couplings. Then we determine the values where the propagators have poles. 

The denominators of the propagators are a polynomial which is of third order in $p^2$. There are two cases: there is a small region in parameter space where the polynomial has three real roots, and for all other values of the couplings there are one real and two complex conjugate roots. In \figurename\ \ref{regionsdiag1} these zones are labelled III and IV respectively. Let us analyze each region separately.

\subsubsection{Three real roots (region III)}

%\begin{figure}\begin{center}
%\includegraphics[width=.5\textwidth]{threems.pdf}
%\caption{The mass-squareds of the massive excitations found in the region where there are three massive poles (region III). \label{threems}}
%\end{center}\end{figure}

Region III is defined by the polynomial in the denominators of \eqref{a3a3}, \eqref{bb}, and \eqref{ba3} having three real roots. This region is sketched in \figurename\ \ref{regionsdiag1}. (Mark that the tip of the region is distorted due to the difficulty in accessing this part numerically.) 

%The square of the masses corresponding to these three roots are plotted in \figurename\ \ref{threems}.

The residues of related to these poles were computed numerically. Only the two of the three roots have positive residue and can correspond to physical states. Those are the one with highest and the one with lowest mass squared. The third of the roots, the one of intermediate value, has negative residue and thus belongs to some negative-norm state, which cannot be physical.

All three states have non-zero mass for non-zero values of the electromagnetic coupling $g'$, with the lightest of the states becoming massless in the limit $g'\to0$. In this limit we recover the behaviour found in this regime in the pure $SU(2)$ case \cite{Capri:2012cr} (the $Z$-boson field having one physical and one negative-norm pole in the propagator) with a massless fermion decoupled from the non-Abelian sector.

\subsubsection{One real root (region IV)}
\begin{figure} \begin{center}
\includegraphics[width=.5\textwidth]{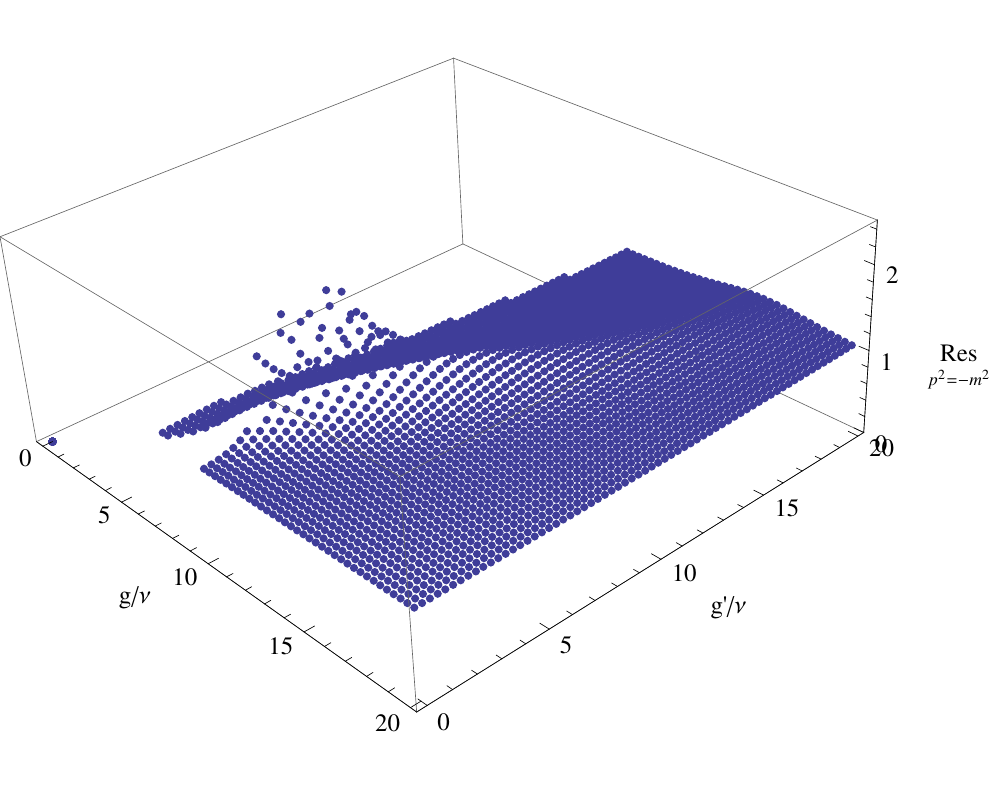}
\caption{The residue of the pole of the photon propagator. It turns out to be positive for all values of the couplings within the region IV. \label{resrealmass}}
\end{center} \end{figure}

In the remaining part of parameter space, there is only one state with real mass-squared. The two other roots of the polynomial in the denominators of \eqref{a3a3}, \eqref{bb}, and \eqref{ba3} have non-zero imaginary part and are complex conjugate to each other. In order to determine whether the pole coming from the real root corresponds to a physical particle excitation, we computed its residue, which can be read off in the partial fraction decomposition (the result is plotted in \figurename\ \ref{resrealmass}). It turns out the residue is always positive, meaning that this excitation has positive norm and can thus be interpreted as a physical, massive particle. The poles coming from the complex roots cannot, of course, correspond to such physical excitations.

%\begin{figure}\begin{center}
%\includegraphics[width=.5\textwidth]{realmass.pdf}
%\caption{The mass-squared of the one physical massive excitation found in region IV. \label{realmass}}
%\end{center}\end{figure}

%\begin{figure}\begin{center}
%\includegraphics[width=.5\textwidth]{realpartccmass.pdf}\includegraphics[width=.5\textwidth]{impartccmass.pdf}
%\caption{The real (left) and imaginary (right) parts of the mass-squared of the other two, complex conjugate, poles. \label{ccmass}}
%\end{center}\end{figure}

In the limit $g'\to0$ we once more recover the corresponding results already found in the pure $SU(2)$ case \cite{Capri:2012cr} (two complex conjugate poles in the propagator of the non-Abelian boson field) plus a massless photon not influenced by the non-Abelian sector.

We shall emphasise here the complexity of the found ``phase spectrum'' in the $3d$ case. For the most part of the $(g'/\nu, g/\nu)$ plane we found the diagonal component of the bosonic field displaying a mix of physical and non-physical contributions, regarding the regions III and IV. The off-diagonal component was found to have physical meaning only in the region I. The transition between those regions was found to be continuous with respect to the effective perturbative parameter $\sim g/\nu$.

\subsection{The $d=4$ case}
\label{d=4}

\begin{figure}\begin{center}
\parbox{.5\textwidth}{\includegraphics[width=.5\textwidth]{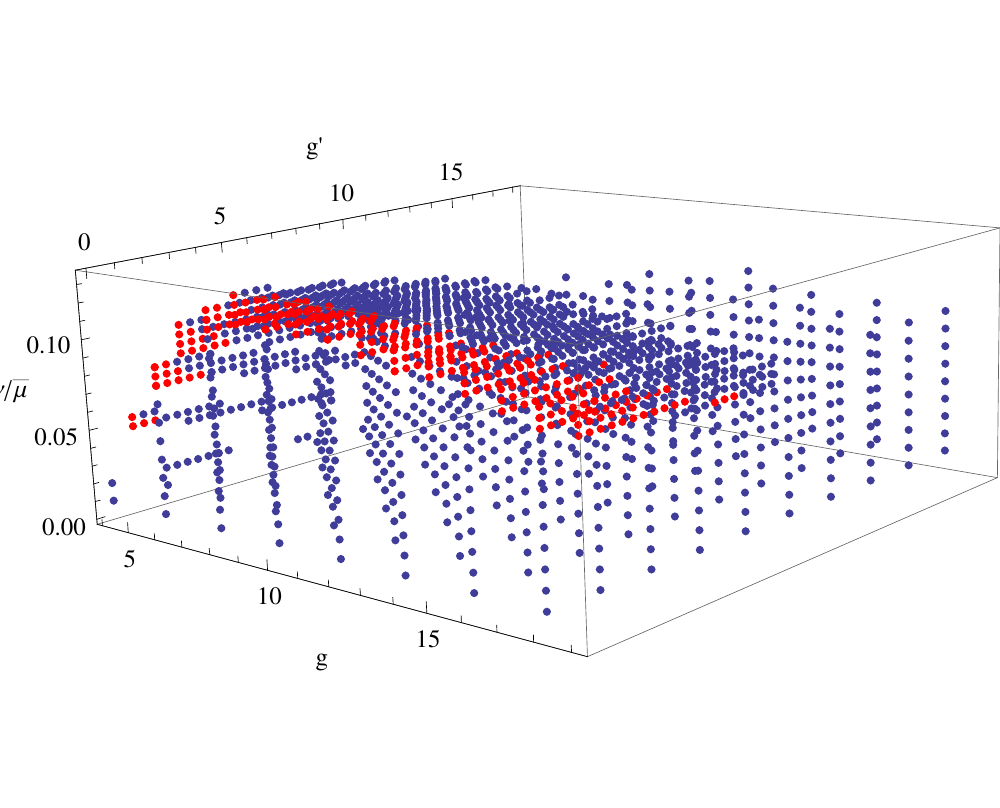}} \quad \parbox{.4\textwidth}{\includegraphics[width=.4\textwidth]{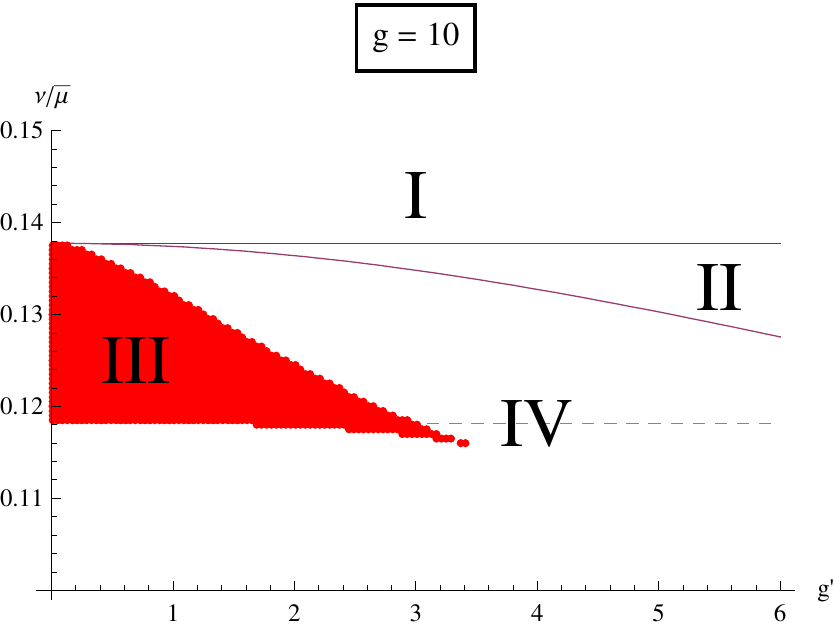}}
\caption{Left is a plot of the region $a'<1/2$ (the region $a'>1/2$ covers all points with higher $\nu$). In red are points where the polynomial in the denominator of \eqref{a3a3} - \eqref{ba3} has three real roots, and in blue are the points where it has one real and two complex conjugate roots. At the right is a slice of the phase diagram for $g=10$. The region $a>1/2$ and $a'>1/2$ is labelled I, the region $a<1/2$ and $a'>1/2$ is II, and the region $a<1/2$ and $a'<1/2$ is split into the regions III (polynomial in the denominator of \eqref{a3a3} - \eqref{ba3} has three real roots, red dots in the diagram at the left) and IV (one real and two complex conjugate roots, blue dots in the diagram at the left). The dashed line separates the different regimes for off-diagonal SU(2) bosons (two real massive poles above the line, two complex conjugate poles below). \label{regionsdiag}}
\end{center}\end{figure}

In the $4$-dimensional case the diagonal and off-diagonal ghost form factors read, using the standard $\MSbar$ renormalization procedure,
\begin{equation}
\label{constas} 
\langle \sigma_\text{off}(0) \rangle = 1 - \frac{3g^{2}}{32\pi^{2}}\ln\frac{2a}{\cos(\theta_{W})}
\,, \qquad
\langle \sigma_\text{diag}(0) \rangle = 1-\frac{3g^{2}}{32\pi^{2}}\ln(2a)\;,
\end{equation}
where
\begin{equation}
\label{consta} a = \frac{\nu^{2}g^{2}}{4\bar{\mu}^{2}\,e^{1-\frac{32 \pi^{2}}{3g^{2}}}}\,, \qquad a' = \frac{\nu^{2}(g^{2}+g'^{2})}{4\bar{\mu}^{2}\,e^{1-\frac{32 \pi^{2}}{3g^{2}}}} = a \frac{g^{2}+g'^{2}}{g^{2}} = \frac{a}{\cos^{2}(\theta_{W})}
\end{equation}
and $\theta_{W}$ stands for the Weinberg angle. With expression \eqref{constas} we are able to identify three possible regions, depicted in \figurename\ \ref{regionsdiag}:
\begin{itemize}
\item Region I, where $\langle \sigma_\text{diag}(0)\rangle < 1$ and $\langle \sigma_\text{off}(0)\rangle < 1$, meaning $2a > 1$. In this case the Gribov parameters are both zero so that we have the massive $W^{\pm}$ and $Z$, and a massless photon. That region can be identified with the ``Higgs phase''.
\item Region II, where $\langle \sigma_\text{diag}(0)\rangle > 1$ and $\langle \sigma_\text{off}(0)\rangle < 1$, or equivalently $\cos \theta_{W} < 2a < 1$. In this region we have $\omega = 0$ while $\beta \neq 0$, leading to a modified $W^{\pm}$ propagator, and a free photon and a massive $Z$ boson.
\item The remaining parts of parameter space, where $\langle \sigma_\text{diag}(0)\rangle > 1$ and $\langle \sigma_\text{off}(0)\rangle > 1$, or $0 < 2a < \cos\theta_{W}$. In this regime we have both $\beta \neq 0$ and $\omega \neq 0$, which modifies the $W^{\pm}$, $Z$ and photon propagators. Furthermore this region will fall apart in two separate regions III and IV due to different behaviour of the propagators (see \figurename\ \ref{regionsdiag}).
\end{itemize}

\subsubsection{The off-diagonal gauge bosons} \label{sect4}
Let us first look at the behaviour of the off-diagonal bosons under the influence of the Gribov horizon. The propagator \eqref{ww} only contains the $\beta$ Gribov parameter, meaning $\omega$ does not need be considered here.

As found in the previous section, this $\beta$ is not necessary in the regime $a>1/2$, due to the ghost form factor $\langle\sigma_\text{diag}(0)\rangle$ always being smaller than one. In this case, the off-diagonal boson propagator is simply of the massive type.
%\begin{equation}
%	\langle W^{+}_{\mu}(p)W^{-}_{\nu}(-p) \rangle = \frac{1}{p^{2} + \frac{\nu^{2}}{2}g^{2}}T_{\mu\nu}(p^2) \;.
%\end{equation}

In the case that $a<1/2$, the relevant ghost form factor is not automatically smaller than one anymore, and the Gribov parameter $\beta$ becomes necessary. The value of $\beta$ is given by the gap equations \eqref{beta gap eq}, which has exactly the same form as in the case without electromagnetic sector. Therefore the results will also be analogous. As the analysis is quite involved, we just quote the results here.

For $1/e<a<1/2$ the off-diagonal boson field has two real massive poles in its two-point function. One of these has a negative residue, however. This means we find one physical massive excitation, and one unphysical mode in this regime. When $a<1/e$ the two poles acquire a non-zero imaginary part and there are no poles with real mass-squared left. In this region the off-diagonal boson propagator is of Gribov type, and the $W$ boson is completely removed from the spectrum. More details can be found in \cite{Capri:2012ah}.

\subsubsection{The diagonal SU(2) boson and the photon} \label{zandgamma}
The two other gauge bosons --- the diagonal SU(2) boson and the photon, $Z_\mu$ and the $A_\mu$ --- have their propagators given by \eqref{zz}, \eqref{aa} and \eqref{az}. Here, $\omega$ is the only of the two Gribov parameters present.

In the regime $a'>1/2$, $\omega$ is not necessary to restrict the region of integration to $\Omega$. Due to this, the propagators are unmodified in comparison to the perturbative case.
%\begin{equation} \label{propsrewrite}
%	\langle Z_{\mu}(p) Z_{\nu}(-p) \rangle = \frac{1}{p^{2} + \frac{\nu^{2}}{2}(g^{2} + g'^{2})} T_{\mu\nu}(p^2)\;, \quad
%	\langle A_{\mu}(p) A_{\nu}(-p) \rangle = \frac{1}{p^{2}}T_{\mu\nu}(p^2) \;.
%\end{equation}

In the region $a'<1/2$ the Gribov parameter $\omega$ does become necessary, and it has to be computed by solving its gap equation. Due to its complexity it seems impossible to compute analytically. Therefore we turn to numerical methods. 
%Using Mathematica the gap equation can be straighforwardly solved for a list of values for the couplings. In order to do this, we regularize the momentum integration by subtracting a term designed to cancel the large-$p^2$ divergence (as in the Pauli--Villars procedure):
%\begin{equation}
%	\frac{3}{2}g^{2}\int \frac{d^{4}p}{(2\pi)^{4}} \left(\frac{p^{2} + \frac{\nu^{2}}{2}g'^{2}}{p^{6} + \frac{\nu^{2}}{2}(g^{2} + g'^{2})p^{4} + \frac{\omega^{\ast}}{2}g^{2}p^{2} + \frac{\omega^{\ast}}{4}\nu^{2}g^{2}g'^{2}} - \frac1{(p^2+M^2)^2}\right) 	+ \frac{3}{2}g^{2}\int \frac{d^{4}p}{(2\pi)^{4}}\frac1{(p^2+M^2)^2} = 1 \;,
%\end{equation}
%where $M^2$ is an arbitrary mass scale. The second integral is readily computed by hand, whereas the first one converges and can be determined numerically. 
Once the parameter $\omega$ has been (numerically) determined, we look at the propagators to investigate the nature of the spectrum.

As the model under consideration depends on three dimensionless parameters ($g$, $g'$ and $\nu/\bar\mu$), it is not possible to plot the parameter dependence of these masses in a visually comprehensible way. Therefore we limit ourselves to discussing the behaviour we observed.

In region III, when there are three real poles in the full two-point function, it turns out that only the two of the three roots we identified have a positive residue and can correspond to physical states, being the one with highest and the one with lowest mass squared. The third one, the root of intermediate value, has negative residue and thus belongs to some negative-norm state, which cannot be physical. All three states have non-zero mass for non-zero values of the electromagnetic coupling $g'$, with the lightest of the states becoming massless in the limit $g'\to0$. In this limit we recover the behaviour found in this regime in the pure $SU(2)$ case \cite{Capri:2012ah} (the $Z$-boson field having one physical and one negative-norm pole in the propagator) with a massless boson decoupled from the non-Abelian sector.

In region IV there is only one state with real mass squared --- the other two having complex mass squared, conjugate to each other --- and from the partial fraction decomposition follows that it has positive residue. This means that, in this region, the diagonal-plus-photon sector contains one physical massive state (becoming massless in the limit $g'\to0$), and two states that can, at best, be interpreted as confined.

%%%%%%%%%%%%%%%%%%%%%%%%%%%%%%%%%%%%%
\section{Conclusion}
\label{conclusion}
%%%%%%%%%%%%%%%%%%%%%%%%%%%%%%%%%%%%%
In this work we presented results achieved during the study of Yang-Mills models, in the Landau gauge, coupled to a Higgs field, taking into account non-perturbative effects. More specifically, the $SU(2)$ and $SU(2)\times U(1)$ models were analysed in $3$-- and $4$--dimensional Euclidean space-time, while the Higgs field was considered in its fundamental and adjoint representation. The non-perturbative effects were taken into account by considering the existence of Gribov ambiguities, or Gribov copies, in the (Landau) gauge fixing process. In order to get rid of those ambiguities we followed the procedure developed by Gribov in his seminal work \cite{Gribov:1977wm}, which consists in restricting the configuration space of the gauge field into the first Gribov region $\Omega$. As found by Gribov, that restriction of the path integral domain leads to a modification of the gluon propagator, in a way that it is not possible to attach any physical particle interpretation to that quantity. The gauge field propagator develops, after the Gribov restriction, two complex conjugate poles, which spoils the physical interpretation of a particle. This may be interpreted as a sign of confinement of the gluon field. In the same sense we could observe similar modifications of the gluon propagator in the probed models. In general, the poles of the gauge field propagator are functions of the parameters in each Yang-Mills model (including the Gribov parameter $\gamma$ and the Higgs self mass parameter $\nu$), so that we could identify regions in the parameter space where the gauge propagator has contributions coming from Yukawa type modes (with real poles) and/or contributions from Gribov like modes (with {\it cc} poles). Regions where only Yukawa contribution exist are called {\it Higgs regime} or {\it Higgs ``phase''}. At the same time, regions where there is only Gribov type contributions are named {\it confined regime} or {\it confined ``phase''}. Note that contributions with negative residue, despite being of the Yukawa type, have no physical particle interpretation as well.

In general we could find that the Higgs regime corresponds to the region of weak coupling, {\it i.e.} small $g$ and sufficiently large $\nu$, reached in the UV regime. For that region of the parameter space (coupling constant and Higgs vacuum expectation value), where we expect perturbation theory to work, we do recover the perturbative Yang-Mills-Higgs propagators. This is an important observation, since it means that the Gribov ambiguity does not spoils the physical vector boson interpretation of the gauge sector, where it is relevant. On the other side, the confined phase corresponds to the strong coupling region in the parameter space, characterized by large values of $g$ and sufficiently small $\nu$, but still keeping logarithmic divergences under control. For higher values of the non-Abelian coupling constant the Gribov horizon lets its influence be felt and the propagators become modified. In general, for the $SU(2)$ and for the $SU(2)\times U(1)$ cases, we could find an intermediate region where contributions from physical modes (with real poles) mix with contributions from non-physical modes (with {\it cc} poles or negative residues). 

For the fundamental representation of the Higgs field, either in the $SU(2)$ or in the $SU(2)\times U(1)$ model, we could find that these two ``phases'' (confinement/Higgs) may be continuously connected. This feature seems to be in qualitative agreement with the work by Fradkin \& Shenker  \cite{Fradkin:1978dv}. According to \cite{Fradkin:1978dv,Caudy:2007sf,Bonati:2009pf} one observes a region in the $(\nu,g)$ plane, called analyticity region, in which the Higgs and confining ``phases'' are connected by paths along which the expectation value of any local correlation function varies analytically, implying the absence of any discontinuity  in the thermodynamical quantities.  According to  \cite{Fradkin:1978dv}, the spectrum of the theory evolves continuously from one regime to the other. It turns out that these two regions are smoothly connected. In our work no rigorous analysis were carried out concerning the existence of any analyticity region, given the obvious complexity of the matter.

The word ``phase'' was carefully used in this work, since the existence of a gauge invariant local operator, whose {\it vev} is sensitive to a phase transition, was not probed in this work. At finite temperature the use of Polyakov loop becomes a useful order parameter, and works on this subject point to second order phase transition for the $SU(2)$ gauge theory (see \cite{Fukushima:2012qa,Marhauser:2008fz} and a more recent work \cite{Canfora:2015yia}).

In the adjoint representation of the Higgs field things look quite different. Besides the confining phase, in which the gluon propagator is of the Gribov type, our results indicate the existence of what can be called a $U(1)$ confining phase for finite values of the Higgs condensate. This is a phase in which the third component $A^3_\mu$ of the gauge field  displays a propagator of the Gribov type, while the remaining off-diagonal components $A^{\alpha}_\mu$, $\alpha=1,2$, exhibit a propagator of the Yukawa type. Interestingly, this phase has been already detected in the lattice studies of the three dimensional Georgi-Glashow model   \cite{Nadkarni:1989na,Hart:1996ac}. A second result is the absence of the Coulomb phase for finite Higgs condensate. For an infinite value of the latter, we were able to clearly reveal the existence of a massless photon, in agreement with the lattice suggestion, as e.g. \cite{Brower:1982yn}.

That non-perturbative analysis of Yang-Mills theories coupled to a Higgs field can be extended in many interesting ways. For instance, a similar analysis could be done considering the all-order approach to the Gribov problem, known as Gribov-Zwanziger approach \cite{Zwanziger:1988jt,Zwanziger:1989mf,Dudal:2009bf} (see also \cite{Sobreiro:2005ec,Vandersickel:2012tz} for detailed study), and also the refined version of it, when non-perturbative effects concerning the existence of dimension two condensates are taken into account \cite{Dudal:2008sp,Dudal:2008rm,Dudal:2010tf,Oliveira:2012eh,Dudal:2012zx}. Equally interesting is the analysis of these Yang-Mills models at finite temperature, which is already being carried out by some of the authors \cite{Canfora:2015yia}.

%===============================================================================

\bibliographystyle{unsrt}

\end{document}